\documentclass[11pt,a4paper]{article}
\usepackage[english]{babel}
\usepackage[utf8]{inputenc}
\usepackage{amsfonts}
\usepackage{amsmath}
\usepackage{amssymb}
\usepackage{mathtools}
\usepackage{graphicx}
\usepackage{bm}
\usepackage{xcolor}
\usepackage[font=small]{caption}
\usepackage{subcaption}
\usepackage{cite}
\def\linkcolor{cyan!70!black}
\usepackage[
colorlinks=true
,urlcolor=\linkcolor
,anchorcolor=\linkcolor
,citecolor=\linkcolor
,filecolor=\linkcolor
,linkcolor=\linkcolor
,menucolor=\linkcolor
,linktocpage=true
,pdfproducer=medialab
,pdfa=true
]{hyperref}
\usepackage{geometry}
\newcommand{\mathsym}[1]{{}}

\DeclareUnicodeCharacter{2212}{-}
\pdfoutput=1
\input{tcilatex}

\DeclareUnicodeCharacter{2212}{-}
\pdfoutput=1
\input{tcilatex}
\begin{document}

\begin{titlepage}
\begin{flushright}
\end{flushright}

\begin{center}
\vspace*{-1cm}
{\Large\bf Three-Loop Inverse Scotogenic Seesaw Models}

\vspace*{0.8cm}

Asmaa Abada$^{a}$\footnote[1]{\href{mailto:asmaa.abada@ijclab.in2p3.fr}{asmaa.abada@ijclab.in2p3.fr}}, Nicolás Bernal$^{b}$\footnote[2]{\href{mailto:nicolas.bernal@nyu.edu}{nicolas.bernal@nyu.edu}}, A. E. Cárcamo Hernández$^{c,d,e}$\footnote[3]{\href{mailto:antonio.carcamo@usm.cl}{antonio.carcamo@usm.cl}}, \\Sergey Kovalenko$^{e,f}$\footnote[4]{\href{mailto:sergey.kovalenko@unab.cl}{sergey.kovalenko@unab.cl}}, Téssio B. de Melo$^{e,f}$\footnote[5]{\href{mailto:tessiomelo@institutosaphir.cl}{tessiomelo@institutosaphir.cl}}\\

\vspace*{0.7cm}
{\small
$^{a}$Pôle Théorie, Laboratoire de Physique des 2 Infinis Irène Joliot Curie (UMR 9012)\\
CNRS/IN2P3,15 Rue Georges Clemenceau, 91400 Orsay, France\\
$^{b}$New York University Abu Dhabi\\
PO Box 129188, Saadiyat Island, Abu Dhabi, United Arab Emirates\\
$^{{c}}$Universidad Técnica Federico Santa María, Casilla 110-V, Valparaíso, Chile\\
$^{{d}}$Centro Científico-Tecnológico de Valparaíso, Casilla 110-V, Valparaíso, Chile\\
$^{{e}}$Millennium Institute for Subatomic Physics at the High-Energy Frontier, SAPHIR, Chile\\
$^{f}$  Universidad Andrés Bello, Facultad de Ciencias Exactas, \\ Departamento de Ciencias Físicas-Center for Theoretical and Experimental Particle Physics, \\
Fernández Concha 700, Santiago, Chile\\
\vspace*{0.5cm}
}

\begin{center}
    {\it Dedicated to the memory of Iván Schmidt, a very nice person,\\
    friend and long-term collaborator who passed away on November 27$^\text{th}$, 2023.}
\end{center}
\vspace*{0.2cm}

\begin{abstract}
    \noindent We propose a class of  models providing an explanation of the origin of light neutrino masses, the baryon asymmetry of the Universe via leptogenesis and offering viable dark matter candidates. In these models the Majorana masses of the active neutrino are generated by the inverse seesaw mechanism with the lepton number violating right-handed Majorana neutrino masses $\mu$ arising at three loops. The latter is ensured by the preserved discrete symmetries, which also guarantee the stability of the dark matter candidate. We focus on one of these models and perform a detailed analysis of the phenomenology of its leptonic sector. The model can successfully accommodate baryogenesis through leptogenesis in both weak and strong washout regimes. The lightest heavy fermion turns out to be a viable dark matter candidate, provided that the entries of the Majorana submatrix $\mu$  are in the keV to MeV range. The solutions are consistent with the experimental constraints, accommodating both mass orderings for active neutrinos, in particular charged-lepton flavor violating decays $\mu\to e\gamma$, $\mu\to eee$, and the electron-muon conversion processes get sizable rates within future sensitivity reach.
\end{abstract}
\end{center}
\end{titlepage}

\section{Introduction}
Several New Physics models beyond the Standard Model (SM) of particle physics have been proposed to accommodate the neutrino oscillation phenomenon and thus to generate masses and mixings for the active neutrinos. The simplest and most elegant extensions of the SM rely on adding new particles while keeping the same gauge symmetry to generate neutrino masses and mixings at tree level: that is, the seesaw mechanism. Among the ingredients, there are right-handed (RH) Majorana neutrinos for the type-I seesaw, scalar isospin doublets for the type-II, and a mixture between the latter two for the type-III seesaw, which necessitates the inclusion of fermion triplets. In most cases, to comply with neutrino data, the additional states are either too heavy to be detected or, if lighter, they couple to the SM via tiny Yukawa couplings. In both cases, the possibilities of testing tree-level neutrino mass generation are very limited, unless the model is enlarged with some extra symmetries (like in the case of minimal flavor violation) and further fields. In addition, it is difficult to comply with the total relic dark matter (DM) density of the Universe with these tree-level scenarios. 

Alternatively, radiative seesaw models are viable and testable extensions of the SM explaining the tiny neutrino masses and their mixings, while the seesaw mediators play an important role in successfully accommodating the observed amount of DM relic density. In most radiative seesaw models, light neutrino masses are generated at the one-loop level. Additionally, to comply with neutrino data, one needs to assume very small neutrino Yukawa couplings, of the order of ${\mathcal{O}}(10^{-6})$, or unnaturally small mass splitting between the CP-even and CP-odd components of the neutral scalar mediators; see \textit{e.g.} Ref.~\cite{Cai:2017jrq} for a review and Refs.~\cite{Jana:2019mgj, Arbelaez:2022ejo} for comprehensive studies of one- and two-loop radiative neutrino mass models. Two-loop neutrino mass models have been proposed in the literature~\cite{Bonilla:2016diq, Baek:2017qos, Saad:2019vjo, Nomura:2019yft, Arbelaez:2019wyz, Saad:2020ihm, Xing:2020ezi, Chen:2020ptg, Nomura:2020dzw} to provide a more natural explanation for the tiny active neutrino masses than those based on the one-loop radiative seesaw.

In this work, we consider neutrino masses generated at the three-loop level~\cite{Krauss:2002px, Aoki:2008av, Kajiyama:2013lja, Ahriche:2014cda, Ahriche:2014oda, Hatanaka:2014tba, Chen:2014ska, Jin:2014glp, Okada:2015hia, Nishiwaki:2015iqa, Ahriche:2015wha, CarcamoHernandez:2015hjp, Gu:2016xno, Cheung:2017efc, Dutta:2018qei, CarcamoHernandez:2019cbd, Cepedello:2020lul, Hernandez:2021iss, Abada:2022dvm, Bonilla:2023wok, Abada:2023znk} to provide a more natural explanation for the smallness of active neutrino masses than those relying on one- or two-loop seesaw realizations. We recall that in the latter constructions, a significant number of new particles has to be considered, small Yukawa couplings are required, and most of the time one type of DM candidate is available, usually fermionic. In our previous work~\cite{Abada:2022dvm, Abada:2023znk}, we proposed an extended inert doublet model, in which light-active neutrino masses arise from a three-loop-level seesaw mechanism, realized by enlarging the SM group with a spontaneously broken global symmetry $U(1)'$ and a preserved $\mathbb{Z}_2$ parity that forbids the generation of neutrino masses at one- and two-loop orders. The scalar sector was also enlarged by including four neutral gauge singlet scalars, whereas the fermion content was augmented with two RH Majorana neutrinos. We showed that this model is consistent with neutrino oscillation data and can successfully accommodate the DM relic abundance while being consistent with bounds arising from charged lepton flavor violation (cLFV) and electroweak precision observables (oblique parameters $S$, $T$, and $U$, in addition to being consistent with the recently observed $W$-mass anomaly~\cite{CDF:2022hxs}).

In the present work, we propose to further explain the origin of the RH Majorana neutrino masses at three-loop order, which, together with a dynamical origin of the lepton number violation (LNV) at the keV-MeV scale, leads to an inverse seesaw (ISS) realization. For this purpose, we investigate which class of three-loop seesaw realization can generate the ISS neutrino texture and consider two topologies of three-loop diagrams, based on which we propose three well-motivated models where Majorana mass submatrices are generated at the three-loop level.

These three models are characteristic examples of a class of models with different symmetries and particle field content that are capable of dynamically generating an ISS mechanism for light and active neutrino masses, with a LNV ($\Delta L = 2$) source parameter $\mu$ generated at three-loop order at the keV-MeV scale. This was inspired by the general formulation of the ISS~\cite{Mohapatra:1986bd}, where the smallness of $\mu$ was attributed to the supersymmetry breaking effects in a $E_6$ scenario inspired by superstrings. In the context of a non-supersymmetric $SO(10)$ model, which contains remnants of a larger $E_6$ symmetry, the other relevant terms of the neutrino mass matrix are generated at two loops, while $\mu$ is generated at higher loops, justifying its smallness~\cite{Ma:2009gu}.

Here, we focus on the potential of one of these three models and conduct a thorough study on the scalar sector, the dynamical generation of the ISS, the viable DM candidates, and their possible phenomenological impact. Regarding the other two example models, we briefly summarize them, leaving their more detailed investigation for future work.

The paper is organized as follows: After a detailed description of Model 1 in Section~\ref{Sec:models}, in Section~\ref{sec:Model3} we study its scalar sector and the neutrino mass generation. Section~\ref{sec:pheno} is devoted to the phenomenological consequences of Model 1, particularly in the violation of the charged-lepton flavor and in the prospects for neutrinoless double-beta decay. Solutions to baryon asymmetry of the Universe (BAU) through leptogenesis and DM problems are discussed in Sections~\ref{lepto} and~\ref{Sec:DM}, respectively. The interplay between LNV, charged-lepton flavor violation, and DM is discussed in Section~\ref{sec:interplay}. Models 2 and 3 are briefly discussed in Section~\ref{sec:Examples}. Finally, the main findings of this work are collected in Section~\ref{Sec:conclusions}.

\section{Model setup} \label{Sec:models}
In this section, we construct a model with scotogenic ISS at three-loop level, which is an extension of the SM with gauge singlet fields: three scalars $\varphi_{1,2}, \sigma$, two left-handed Majorana neutrinos $\Omega_{1,2}$ and two vector-like neutral leptons $\Psi_{1,2}$. The SM gauge symmetry is extended with the global symmetry $U(1)' \otimes \mathbb{Z}_2$.  As will be discussed in the next section, the model scalar potential develops tree-level instability forming vacuum expectation values (VEVs) of $\langle\phi\rangle = v_\phi, \langle\sigma\rangle = v_\sigma$ that break the symmetry according to
{\Large 
\begin{center}
    \begin{table}[t]\centering
{\renewcommand{\arraystretch}{1.0}  
\begin{tabular}{|c|c|c|c|c|c|c|c|c|c|c|c|}
\hline
Field & $l_{iL}$ & $l_{iR}$ & $\nu _{kR}$ & $N_{kR}$ & $\Psi _{kR}$ & $\Psi
_{kL}$ & $\Omega _{kL}$ & $\phi $ & $\varphi _{1}$ & $\varphi _{2}$ & $%
\sigma $ \\ \hline\hline
$SU(3)_{C}$ & $\mathbf{1}$ & $\mathbf{1}$ & $\mathbf{1}$ & $\mathbf{1}$ & $%
\mathbf{1}$ & $\mathbf{1}$ & $\mathbf{1}$ & $\mathbf{1}$ & $\mathbf{1}$ & $%
\mathbf{1}$ & $\mathbf{1}$ \\ \hline
$SU(2)_{L}$ & $\mathbf{2}$ & $\mathbf{1}$ & $\mathbf{1}$ & $\mathbf{1}$ & $%
\mathbf{1}$ & $\mathbf{1}$ & $\mathbf{1}$ & $\mathbf{2}$ & $\mathbf{1}$ & $%
\mathbf{1}$ & $\mathbf{1}$ \\ \hline
$U(1)_{Y}$ & $-\frac{1}{2}$ & $-1$ & $0$ & $0$ & $0$ & $0$ & $0$ & $\frac{1}{%
2}$ & $0$ & $0$ & $0$ \\ \hline
$U(1)^{\prime }$ & $0$ & $4$ & $-4$ & $4$ & $-5$ & $-1$ & $0$ & $-4$ & $-1$
& $-1$ & $4$ \\ \hline
$\mathbb{Z}_{2}$ & $0$ & $0$ & $0$ & $0$ & $1$ & $1$ & $1$ & $0$ & $1$ & $0$
& $0$ \\ \hline\hline
$\mathbb{Z}_{4}$ & $0$ & $0$ & $0$ & $0$ & $-1$ & $-1$ & $0$ & $0$ & $-1$ & $%
-1$ & $0$ \\ \hline
\end{tabular}
}
\caption{Model charge assignments under the $SU(3)_{C}\otimes
SU(2)_{L}\otimes U(1)_{Y}\otimes U(1)^{\prime }\otimes \mathbb{Z}_{2}$
symmetry and the residual $\mathbb{Z}_{4}$. Here $i=1$, 2, 3 and $k=1$, 2.}
\label{model1}
\end{table}
\end{center}}
\begin{align}
    \label{eq:Symmetry-1}
    &SU(3)_{C}\otimes SU(2)_L\otimes U(1)_Y\otimes U(1)' \otimes \mathbb{Z}_2 \\\nonumber
    &\hspace{30mm}\Downarrow v_\sigma  \\
    \label{eq:Symmetry-2}
    &SU(3)_{C}\otimes SU(2)_L\otimes U(1)_Y \otimes \mathbb{Z}_4 \otimes \mathbb{Z}_2 \\\nonumber
    &\hspace{30mm}\Downarrow v_{\phi}   \\
    \label{eq:Symmetry-3}
    &SU(3)_{C}\otimes U(1)_\text{em} \otimes \mathbb{Z}_4 \otimes \mathbb{Z}_2\,,
\end{align}
where $\phi$ corresponds to the SM Higgs doublet. The global $U(1)'$ symmetry is spontaneously broken at the TeV scale by the VEV of $\langle\sigma\rangle$ down to a residual preserved $\mathbb{Z}_4$ symmetry. This results in the new particles acquiring masses on the TeV scale and hence inducing interesting phenomenological consequences, as discussed in the next sections. Other new scalars with nontrivial $\mathbb{Z}_2$ charges do not acquire VEVs to maintain this symmetry unbroken. The assignment of charge for the leptonic fields for the extended symmetry in Eq.~\eqref{eq:Symmetry-1} is shown in Table~\ref{model1}. We do not show the quark field assignments since in the present work we focus on the lepton sector. Furthermore, it is worth mentioning that spontaneous breaking of the global $U(1)'$ symmetry will give rise to a Goldstone boson corresponding to the imaginary part of the  gauge singlet scalar field $\sigma$. Assuming that the global $U(1)'$ symmetry is also softly broken, such a Goldstone boson will acquire a mass via a $U(1)'$ soft-breaking mass term $\mu_{sb}^2\left(\sigma^2 + \text{H.c.}\right)$ in the scalar potential. Note also that this term does not generate other soft breaking terms radiatively and, therefore, its introduction does not change other aspects of the model except for engendering a mass to the $U(1)'$ Goldstone boson. Alternatively, the Goldstone boson can also acquire a mass from non-perturbative QCD effects associated with the mixed $\left(SU(3)_C\right)^2U(1)'$ anomaly, in an extended version of our model where the quark fields will have non-trivial $U(1)'$ assignments. As discussed in detail in Ref.~\cite{Barger:2008jx}, the Goldstone boson arising from the spontaneous breaking of the global $U(1)'$ symmetry can be a scalar DM candidate for appropriate values of its mass and couplings and can play a crucial role for Electroweak Phase Transition. Here, we do not consider a detailed analysis of that scenario, which deserves another study, which is left beyond the scope of our work.

\begin{figure}[t]
    \begin{center}
        \includegraphics[width=0.65\textwidth]{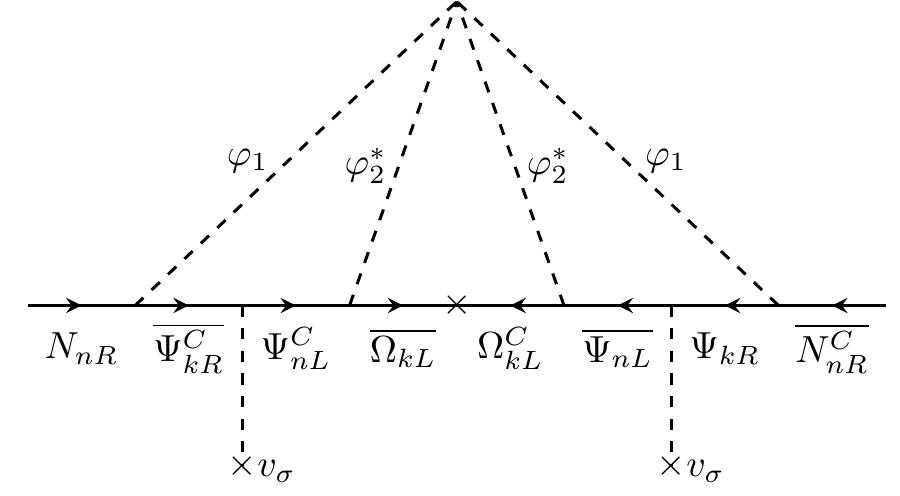}
    \end{center}
    \caption{Three-loop diagram for the LNV mass term in model 1, with $n, k = 1, 2$.}
    \label{Neutrinodiagram1}
\end{figure}

The Yukawa interactions relevant to the neutrino mass generation in our model  are given by 
\begin{align} \label{Eq:Lag3}
    -\mathcal{L}_Y^{\left(\nu\right)} =& \sum_{i=1}^3 \sum_{k=1}^2 \left(y_{\nu}\right)_{ik}\, \overline{l}_{iL}\, \widetilde{\phi}\, \nu_{kR} + \sum_{n=1}^2 \sum_{k=1}^2\, M_{nk}\, \overline{\nu}_{nR}\, N_{kR}^C \notag \\
    &+ \sum_{n=1}^2 \sum_{k=1}^2 \left(y_{N}\right)_{nk}N_{nR}\, \varphi_1^*\, \overline{\Psi_{kR}^C} + \sum_{n=1}^2 \sum_{k=1}^2 \left(y_{\Omega}\right)_{nk}\, \Psi_{nL}^C\, \varphi_2\overline{\Omega}_{kL} \notag \\
    &+ \sum_{n=1}^2 \sum_{k=1}^2 \left(y_{\Psi}\right)_{nk}\, \overline{\Psi}_{nL}\, \sigma\, \Psi_{kR} + \sum_{n=1}^2 \sum_{k=1}^2 \left(m_{\Omega}\right)_{nk}\, \overline{\Omega}_{kL}\, \Omega_{nL}^C + \text{H.c.}
\end{align}
Scalar fields $\varphi_1$ and $\varphi_2$ do not acquire vacuum expectation values, but together with the heavy neutral leptons $\Psi_{kR}$, $\Psi_{kL}$ and $\Omega_{kL}$ (with $k=1$, 2) induce the LNV Majorana mass term $\mu_{nk}\, \overline{N_{nR}}\, N_{kR}^C$ at the three-loop level according to the diagram shown in Fig.~\ref{Neutrinodiagram1}. Note that we assign lepton numbers $L$ to the fields $L(l_L)=-1$, $L(\nu_R)=-1$, $L(N_R)=1$, $L(\Psi_{L,R}) =-1$, and $L(\Omega_L)=1$, so that the accidental $U(1)_L$ lepton number symmetry is softly broken by the mass term $m_\Omega\, \overline{\Omega_{L}}\, \Omega_{L}^C$. Therefore, the tiny active neutrino masses are protected in our model by this accidental $U(1)_L$ symmetry and are technically natural.

It is worth mentioning that the case of only one generation of neutral leptons $\nu _R$, $N _R$, $\Psi _{L,R}$, $\Omega_{L}$ would imply that the entries of the light active neutrino mass matrix arising from the inverse seesaw will be of the form $\left(\widetilde{\mathbf{M}}_{\nu}\right)_{ij}=z_iz_j$ ($i,j=1,2,3$). Such a low-energy neutrino mass matrix will only have one nonvanishing eigenvalue, thus implying that only one active neutrino will acquire a mass, which is in contradiction with the experimental data on neutrino oscillations.

As requested by our strategy, the $U(1)'\otimes \mathbb{Z}_2$ symmetry, as well as the $\mathbb{Z}_4$ symmetry arising from the spontaneous breaking of the $U(1)'$, forbid tree, one-loop and two-loop-level mass generation for active neutrinos, while allowing the three-loop contribution in Fig.~\ref{Neutrinodiagram1}.

In the next sections, the phenomenology of this model will be studied in detail, in particular the scalar potential, the generation of neutrino masses, the DM phenomenology and the dynamical generation of the BAU through leptogenesis.

\section{The scalar potential and generation of neutrino masses} \label{sec:Model3}
Hereinafter, the phenomenology of the described model will be studied in detail.
The most general scalar potential invariant under the symmetry in Eq.~\eqref{eq:Symmetry-1} reads 
\begin{align}
    V=& -\mu_{\phi }^2 (\phi^\dagger \phi )-\mu_\sigma^2 (\sigma^*\sigma )+\mu_{\varphi_1}^2 (\varphi_1^*\varphi_1)+\mu_{\varphi_2}^2 (\varphi_2^*\varphi_2) +\lambda_1(\phi^\dagger \phi )^2 +\lambda_2(\sigma^*\sigma )^2 +\lambda_3(\varphi_1^*\varphi_1)^2   \nonumber\\
    &+\lambda_4(\varphi_2^*\varphi_2)^2 +\lambda_5(\phi^\dagger \phi )(\sigma^*\sigma )+\lambda_6(\phi^\dagger \phi )(\varphi_1^*\varphi_1)+\lambda_7(\phi^\dagger \phi )(\varphi_2^*\varphi_2) +\lambda_8(\sigma^*\sigma )(\varphi_1^*\varphi_1) \notag \\
    \label{eq:V-1}
    &+\lambda_9(\sigma^*\sigma )(\varphi_2^*\varphi_2)+\lambda_{10}(\varphi_1^*\varphi_1)(\varphi_2^*\varphi_2) +\lambda_{11}\left[ \varphi_1^2 (\varphi_2^*)^2 + \text{H.c.}\right], 
\end{align}
where the coefficients $\mu_i$ have mass dimension, while the quartic couplings $\lambda_i$ are dimensionless. The scalar fields of the model can be expanded as 
\begin{equation}
    \phi = \left(
    \begin{array}{c}
        \phi^{+} \\ 
        \frac{1}{\sqrt{2}}\left(v_{\phi}+\phi_R^0+i\,\phi_{I}^0\right) 
    \end{array}\right), \hspace{0.5cm}
    \sigma =\frac{1}{\sqrt{2}}\left(v_\sigma+\sigma_R+i\,\sigma_{I}\right), \hspace{0.5cm}
    \varphi_k =\frac{1}{\sqrt{2}}\left(\varphi_{kR}+i\,\varphi_{kI}\right),
\end{equation}
with $k = 1$, 2.
Here, $\phi^{\pm}$ and $\phi_{I}^0$ are the SM would-be-Goldstone bosons, while $\sigma_{I}$ is a massless physical CP-odd  Goldstone boson, sterile with respect to the SM interactions. Due to the preserved $\mathbb{Z}_2$ symmetry, the neutral CP-even components $\phi_R^0$ and $\sigma_R$ do not mix with the remaining neutral CP-even scalar fields $\varphi_{kR}$. The squared mass matrix for the neutral CP-even scalar fields that transform trivially under the preserved $\mathbb{Z}_2 \otimes \mathbb{Z}_4$ symmetry, on the basis $\left(\phi_R^0,\sigma_R\right)$ can be written as
\begin{equation}
    M_H^2 =\left(
    \begin{array}{cc}
        2\lambda_1\, v_\phi^2  & \lambda_5\, v_\phi\, v_\sigma \\ 
        \lambda_5\, v_\phi\, v_\sigma & 2\lambda_2\,v_\sigma^2 
    \end{array}
    \right),
\end{equation}
where in the decoupling limit $\lambda_5\to 0$, $\phi_R^0$ corresponds to the SM-like 125~GeV Higgs boson. The squared masses for the inert CP-even scalar fields $\varphi_{1R}$, $\varphi_{2R}$ and for the CP-odd scalars $\varphi_{1I}$, $\varphi_{2I}$ are given by
\begin{align}
    m_{\varphi_{1R}}^2 & =\mu_{\varphi_1}^2 +\frac12 \lambda_8\, v_\sigma^2 +\frac12 \lambda_6\, v_\phi^2\, ,  \\
    m_{\varphi_{2R}}^2 & =\mu_{\varphi_2}^2 +\frac12 \lambda_9\, v_\sigma^2 +\frac12 \lambda_7\, v_\phi^2\, , \\
    m_{\varphi_{1I}}^2 & =\mu_{\varphi_1}^2 +\frac12 \lambda_8\, v_\sigma^2 +\frac12 \lambda_6\, v_\phi^2\, , \\
    m_{\varphi_{2I}}^2 & =\mu_{\varphi_2}^2 +\frac12 \lambda_9\, v_\sigma^2 +\frac12 \lambda_7\, v_\phi^2\, .
\end{align}
Finally, the conditions for minimization of the scalar potential yield the following relations 
\begin{align}
    \mu_{\phi }^2 & =\lambda_1\, v_\phi^2 +\frac12 \,\lambda_5\,v_\sigma^2 \,, \\
    \mu_\sigma^2 & =\frac{\lambda_5\, v_\phi^2 }{2}+\lambda_2\,v_\sigma^2 \,.
\end{align}

\subsection{Stability of the scalar potential}
In what follows, we derive tree-level stability conditions that the scalar potential must satisfy in order to be bounded from below. To this end, it is sufficient to analyze the quartic interactions because they dominate the behavior of the scalar potential in the region of very large values of the field components. We introduce the following Hermitian bilinear combinations of the scalar fields 
\begin{align}
    a &=\phi^\dagger \phi ,\hspace{1cm} b = \sigma^* \sigma, \hspace{1cm} c = \varphi_1^* \varphi_1, \hspace{1cm} d = \varphi_{2}^* \varphi_{2},  \notag \\
    e &= \varphi_1 \varphi_{2}^*+\varphi_1^*\varphi_2, \hspace{1cm} f = I \left(\varphi_1\varphi_{2}^*-\varphi_1^*\varphi_2\right) ,
\end{align}
which allow to express the quartic couplings in the form
\begin{equation}
    V_4 = \lambda_1a^2 + \lambda_2b^2 + \lambda_3c^2 + \lambda_4d^2 + \lambda_5ab + \lambda_6ac + \lambda_7ad + \lambda_8bc + \lambda_9bd + \lambda_{10}cd + \frac{\lambda_{11}}{2}\left(e^2-f^2\right),
\end{equation}
or, equivalently, to
\begin{align}
    V_4=& \left(\sqrt{\lambda_1}a-\sqrt{\lambda_2}b\right)^2+\left(\sqrt{\lambda_1}a-\sqrt{\lambda_3}c\right)^2+\left(\sqrt{\lambda_1}a-\sqrt{\lambda_4}d\right)^2 +\left(\sqrt{\lambda_2}b-\sqrt{\lambda_3}c\right)^2 \nonumber\\
    &+\left(\sqrt{\lambda_2}b-\sqrt{\lambda_4}d\right)^2+\left(\sqrt{\lambda_3}c-\sqrt{\lambda_4}d\right)^2  +\left(\lambda_5+2\sqrt{\lambda_1\lambda_2}\right) ab+\left(\lambda_6+2\sqrt{\lambda_1\lambda_3}\right) ac\nonumber\\
    &+\left(\lambda_7+2\sqrt{\lambda_1\lambda_4}\right) ad +\left(\lambda_8+2\sqrt{\lambda_2\lambda_3}\right) bc+\left(\lambda_9+2\sqrt{\lambda_2\lambda_4}\right) bd+\left(\lambda_{10}+2\sqrt{\lambda_3\lambda_4}\right) cd \nonumber\\
    & -2\left(\lambda_1a^2+\lambda_2b^2+\lambda_3c^2+\lambda_4d^2\right) +\frac12\lambda_{11}\left(e^2-f^2\right).
\end{align}
Following the procedure used for analyzing the stability described in Refs.~\cite{Maniatis:2006fs, Bhattacharyya:2015nca, Abada:2021yot, Hernandez:2021kju}, we find that the scalar potential is stable when the following conditions are fulfilled:
\begin{align}
    &\lambda_1 \geq 0,\hspace{1cm}\lambda_2\geq 0,\hspace{1cm}\lambda_3\geq 0,\hspace{1cm}\lambda_4\geq 0,\hspace{1cm}\lambda_{11}\geq 0, \notag \\
    &\lambda_5+2\sqrt{\lambda_1\lambda_2} \geq 0,\hspace{1cm}\lambda_6+2\sqrt{\lambda_1\lambda_3}\geq 0,\hspace{1cm}\lambda_7+2\sqrt{\lambda_1\lambda_4}\geq 0,  \notag \\
    &\lambda_8+2\sqrt{\lambda_2\lambda_3} \geq 0,\hspace{1cm}\lambda_9+2\sqrt{\lambda_2\lambda_4}\geq 0,\hspace{1cm}\lambda_{10}+2\sqrt{\lambda_3\lambda_4}\geq 0.
\end{align}
These conditions are imposed in the numerical analysis presented in the following sections.

\subsection{Generation of neutrino masses and lepton mixing}
The neutrino Yukawa terms in Eq.~\eqref{Eq:Lag3} give rise to the following neutrino mass terms
\begin{eqnarray}
    -\mathcal{L}_\text{mass}^{\left(\nu \right) }&=&\frac12 \left(
    \begin{array}{ccc}
        \overline{\nu_L^C} & \overline{\nu_R} & \overline{N_R}%
    \end{array}
    \right) M_\nu \left(
    \begin{array}{c}
        \nu_L \\ 
        \nu_R^C \\ 
        N_R^C%
    \end{array}
    \right) + \sum_{n=1}^2 \sum_{k=1}^2 \left(m_{\Psi}\right)_{nk}\, \overline{\Psi}_{nL}\Psi_{kR}\notag\\ 
    &+& \sum_{n=1}^2 \sum_{k=1}^2 \left(m_{\Omega}\right)_{nk}\, \overline{\Omega}_{kL}\, \Omega_{nL}^C+H.c ,
    \label{neutrinomassterms}
\end{eqnarray}
where $\left(m_{\Psi}\right)_{nk}=\left(y_{\Psi}\right)_{nk}\frac{v_{\sigma}}{\sqrt{2}}$ the total neutrino mass matrix is written in the basis $\left(\nu_L, \nu_R^C, N_R^C\right)^T$ as
\begin{equation}
    M_\nu =\left(
    \begin{array}{ccc}
        0_{3\times 3} & m_{\nu D} & 0_{3\times 2} \\ 
        m_{\nu D}^{T} & 0_{2\times 2} & M \\ 
        0_{2\times 3} & M^{T} & \mu 
    \end{array}
    \right).
    \label{Mnu}
\end{equation}
While the submatrices $m_{\nu D}$ and $M$ are generated at tree level, the entries of the $2\times 2$ submatrix $\mu$, which violate the total lepton number, are generated at three-loop level, as can be seen in Fig.~\ref{Neutrinodiagram1}.

The entries of the $3\times 2$ Dirac submatrix $m_{\nu D}$ are given by
\begin{equation}
    \left(m_{\nu D}\right)_{ik}  = \left(y_\nu \right)_{ik}\frac{v_\phi}{\sqrt{2}}\,,
\end{equation}
with $i=1$, 2, 3 and $k = 1$, 2. The entries of the $2\times 2$ matrix $\mu$ induced at three-loop are
\begin{equation}
    \mu_{sp} = \sum _{k = 1} ^2 \sum _{n = 1} ^2 \sum _{r = 1} ^2 \frac{\lambda_{10}m_{\Psi_R}m_{\Psi_k}}{8\left(4\pi^2 \right)^3 m_{\varphi_2}}\left(y_{N}\right)_{sr}\left(y_{N}\right)_{pk}\left(y_{\Omega }\right)_{rn}\left(y_{\Omega }\right)_{kn}F\left(\frac{m_{\Omega_{n}}^2 }{m_{\varphi_2}^2 },\frac{m_{\varphi_1}^2 }{m_{\varphi_2}^2 }\right),
\end{equation}
with $n$, $k$, $r$, $s$, $p = 1$, 2, in term of the 3-loop function~\cite{Krauss:2002px, Ahriche:2013zwa, CepedelloPerez:2021tgj}
\begin{equation}
    F(\alpha ,\beta) \equiv \frac{\sqrt{\alpha }}{\beta^2 } \int_{0}^{\infty }dr\frac{r}{r+\alpha }\left(\int_{0}^{1}dx\ln \left[ \frac{x\left(1-x\right) r+\left(1-x\right) \beta +x}{x\left(1-x\right) r+x}\right] \right)^2.
\end{equation}

As shown in detail in Ref.~\cite{Catano:2012kw}, the full rotation matrix
that diagonalizes a neutrino mass matrix of the form of Eq.~\eqref{Mnu} is
given by: 
\begin{equation}
    \mathbb{U} \simeq 
    \begin{pmatrix}
        \left(1_{3\times 3}-\frac{R_1R_1^\dagger }{2}\right) \left(1_{3\times 3}-\frac{R_2R_2^\dagger }{2}\right) R_\nu  & R_1R_{M}^{(1) } & R_2R_{M}^{(2)} \\ 
        -\frac{R_1^\dagger +R_2^\dagger }{\sqrt{2}}R_\nu  & \left(1_{3\times 3}-\frac{R_1R_1^\dagger }{2}\right) \frac{1-S}{\sqrt{2}} R_{M}^{(1) } & \frac{1+S}{\sqrt{2}}R_{M}^{(2)} \\ 
        -\frac{R_1^\dagger -R_2^\dagger }{\sqrt{2}}R_\nu  & -\frac{1+S}{\sqrt{2}}R_{M}^{(1) } & \left(1_{3\times 3} - \frac{R_2R_2^\dagger }{2}\right) \frac{1-S}{\sqrt{2}}R_{M}^{(2)}%
    \end{pmatrix} \label{U}
\end{equation}
where 
\begin{align}
    S &= -\frac14M^{-1}\mu\,, \label{s_mat_def}\\
    R_1 &\simeq R_2\simeq \frac{1}{\sqrt{2}}m_{\nu D}^*M^{-1}=\frac{1}{\sqrt{2}m_{N}}m_{\nu D}^*\, \equiv R.  \label{r_mat_def} 
\end{align}
It is worth mentioning here that the whole $7\times 7 $ matrix $\mathbb{U}$ is unitary. In the above matrix, $R_\nu $, $R_{M}^{(-)}$ and $R_{M}^{(+) }$ are rotation matrices that diagonalize the physical mass matrices, $\widetilde{\mathbf{M}}_\nu $, $\mathbf{M}_\nu ^{(-) }$ and $\mathbf{M}_\nu ^{(+) }$, respectively. Regarding the latter, the obtained mass spectrum exhibits the typical spectrum pattern of the ISS mechanism: light masses corresponding to the active neutrino states and quasi-degenerate masses corresponding to the heavy, mostly sterile states forming pseudo-Dirac neutrino pairs, in which the mass gap is proportional to entries of the lepton number violating mass matrix $\mu$. This can be seen in the following mass matrix eigenstates obtained from the diagonalization~\cite{Casas:2001sr, Das:2012ze, Dolan:2018qpy, Cordero-Carrion:2018xre}
\begin{equation}
    \widetilde{\mathbf{M}}_\nu  = m_{\nu D} \left(M^{T}\right)^{-1} \mu\, M^{-1} m_{\nu D}^{T}, \hspace{0.3cm} \mathbf{M}_{\nu}^{(-)} = -\frac{M+M^{T}}{2} + \frac{\mu}{2}, \hspace{0.3cm} \mathbf{M}_{\nu}^{(+)} = \frac{M+M^{T}}{2} + \frac{\mu}{2}\,,
    \label{M1nu}
\end{equation}
where $\widetilde{\mathbf{M}}_\nu $ corresponds to the (mostly) active neutrino mass matrix, while $\mathbf{M}_{\nu}^{(-) }$ and $\mathbf{M}_{\nu}^{(+) }$ are the exotic neutrino mass matrices. Note that the physical neutrino spectrum is composed of three light active neutrinos and four exotic neutrinos, forming two pairs of pseudo-Dirac neutrino states, with masses $\sim \pm \frac12 \left(M+M^{T}\right) $ and a small splitting $\mu $ in each pair. Also note that the rotation matrix $R_\nu $ diagonalizing the light neutrino mass one, $\widetilde{\mathbf{M}}_\nu $, corresponds to the $U_{\text{PMNS}}$ lepton mixing matrix, which would deviate from unitarity, given the presence of heavy exotic neutrinos mixing with the active ones with the mixings in $R_1$ and $R_2$.

Furthermore, using Eq.~\eqref{U} we find that the neutrino fields $\nu_L = \left(\nu_{1L}, \nu_{2L}, \nu_{3L}\right)^{T}$, $\nu_R^C = \left(\nu_{1R}^C,\nu_{2R}^C\right)^T$ and $N_R^C = \left(N_{1R}^C, N_{2R}^C\right)^T$ are related with the neutrino mass eigenstates by the following relations
\begin{equation}
    \left(
    \begin{array}{c}
        \nu_L \\ 
        \nu_R^C \\ 
        N_R^C%
    \end{array}%
    \right) = \mathbb{U\, N}_L\ , \hspace{0.5cm} {\text{where}} \ \mathbb{N}_L = \left(
    \begin{array}{c}
        \widetilde{\nu }_L \\ 
        N_L^{(-) } \\ 
        N_L^{(+) }
    \end{array}
    \right),
\end{equation}
where the total mixing matrix $\mathbb{UN}_L$ is given by Eq.~\eqref{U}, $\widetilde{\nu}_{jL}$ ($j=1,2,3$), $N_{kL}^{(-) }$ and $N_{kL}^{(+) }$ ($k=1,2$) are the three active neutrinos and four exotic neutrinos, respectively. 

In order to successfully reproduce the neutrino oscillation experimental data, the Dirac submatrix $m_{\nu D}$, in the basis of diagonal SM charged lepton mass matrix, should have the following form~\cite{Casas:2001sr, Ibarra:2003up, Cordero-Carrion:2019qtu, CarcamoHernandez:2019lhv, Hernandez:2021xet}
\begin{equation}
    m_{\nu D}=R_{\nu}\left(\left(\widetilde{\mathbf{M}}_\nu \right)_\text{diag}\right)^{\frac12 }O\mu^{-\frac12 } M\,,
\label{mnuD}
\end{equation}
where
\begin{equation} \label{eq:Mdiag}
    \left(\widetilde{\mathbf{M}}_\nu \right)_\text{diag} = \text{diag}\left(m_1,m_2,m_3\right) 
\end{equation}
being $m_1$, $m_2$ and $m_3$ the masses of the light active neutrinos, $R_{\nu}$ the $U_{\text{PMNS}}$ leptonic mixing matrix (without unitarity violations) and $O$ a complex orthogonal rotation matrix.

\subsection{Neutrinoless double beta decay}
In our model, the only source of LNV is the Majorana neutrino mass terms. Consequently, neutrinoless double beta decay $0\nu\beta\beta$ is mediated by the well-known neutrino mass mechanism. In this case,  its amplitude is  proportional to the effective mass parameter~\cite{Kovalenko:2009td, Faessler:2014kka} 
\begin{equation} \label{mee3p2}
    m_{ee} = \bigg|\sum_{i=1}^N \mathbb{U}_{ei}^2\, p^2 \dfrac{m_i}{p^2+m_i^2}\bigg|, 
\end{equation}
written for $N = 7$ Majorana neutrino mass states existing in the model. Here, $p^2$ is usually interpreted as the mean square of the Fermi momentum of the nucleon for the decaying nucleus. Its value depends on the nucleus and the nuclear structure model used to calculate it. Here, we take the average value for the experimentally interesting isotopes $p^2 \simeq -(150~\mathrm{MeV})^2$~\cite{Faessler:2014kka, Babic:2018ikc}.

The physical neutrino states in our model are light (active) neutrinos and pairs of heavy pseudo-Dirac neutrinos with opposite CP parity. Due to the last fact, the contribution of the pseudo-Dirac pairs to the effective mass in Eq.~\eqref{mee3p2} almost completely annihilates.  The remaining contribution comes from light neutrinos. Thus, we have
\begin{equation}\label{meefp2}
    m_{ee} =\left|\sum_{i=1}^3 \mathbb{U}_{ei}^2\, p^2 \dfrac{m_i}{p^2+m_i^2}\right| \equiv \left|\sum_{i=1}^3 \mathbb{U}_{ei}^2\, m_i\right|,
\end{equation}
which we use in our analysis.

\section{Phenomenology} \label{sec:pheno}

\subsection{Charged lepton flavor violation} \label{sec:clfv}
In this section, we will discuss the implications of our model in the lepton flavor violating decays such as $\ell_\alpha\to \ell_\beta\ell_\beta\ell_\rho$, $\text{CR}(\mu -e, \text{ N})$, and the radiative cLFV decays:  $\mu \to e\gamma$, $\tau \to \mu \gamma$ and $\tau \to e\gamma $. As mentioned in the previous section, the sterile neutrino spectrum of Model~1 is composed of four TeV-scale neutrinos that are practically degenerate. These heavy sterile neutrinos mix with the active ones, with mixing angles of the order of $\frac{1}{\sqrt{2}m_{N}}\left(m_{\nu D}\right)_{in}$ with $i=1$, 2, 3 and $n=1$, 2. The admixture of the heavy sterile neutrinos in the left-handed charged current $SU(2)_L \otimes U(1)_Y$ weak interaction gives rise to the decay $l_i \to l_j \gamma$ at one-loop level, with a branching ratio given by~\cite{Langacker:1988up, Lavoura:2003xp, Hue:2017lak, CarcamoHernandez:2020pnh, Bonilla:2023egs, Bonilla:2023wok, CarcamoHernandez:2023atk, Batra:2023mds}
\begin{align}
    \text{Br}\left(l_{i}\to l_{j}\gamma \right) &= \frac{\alpha_{W}^3 s_{W}^2 m_{l_{i}}^{5}}{256\pi^2 m_{W}^{4}\Gamma_{i}}\left\vert G_{ij}\right\vert^2,\\
    G_{ij} &\simeq \sum_{k=1}^3 \left(\left[\left(1-RR^\dagger \right) R_\nu \right]^*\right)_{ik}\left(\left(1-RR^\dagger \right) R_\nu \right)_{jk}G_{\gamma }\left(\frac{m_{\nu_k}^2 }{m_{W}^2 }\right) \nonumber\\
    &\qquad + 2\sum_{k=1}^2 \left(R^*\right)_{ik}\left(R\right)_{jk}G_{\gamma }\left(\frac{m_{N_k}^2 }{m_{W}^2 }\right), \\
    G_{\gamma}(x) &\equiv \frac{10-43x+78x^2 -49x^3 +18x^3 \ln x+4x^{4}}{12\left(1-x\right)^{4}}, \label{Brmutoegamma2}
\end{align}
where $\Gamma_{\mu } \simeq 3\times 10^{-19}$~GeV is the total muon decay width and $R$ is given in Eq. \eqref{r_mat_def}. Furthermore, to successfully reproduce the experimental neutrino oscillation data, the Dirac submatrix $m_{\nu D}$, in the basis of the diagonal SM charged lepton mass matrix, should have the form given in Eq.~\eqref{mnuD}.

\subsubsection{\boldmath $\text{CR}(\mu -e, \text{N})$}
The $\mu - e$ conversion occurs in a muonic atom formed when a muon is captured, falling into the first state of a target nucleus $N$. The conversion rate is defined as
\begin{equation} \label{eq:CR:def}
    \text{CR}(\mu -e, \text{ N}) \equiv \frac{\Gamma (\mu^- + N \to e^- +N)}{\Gamma (\mu^- + N \to \text{ all)}}\,.
\end{equation}
Box and penguin diagrams contribute as
\begin{equation} \label{eq:conv}
    \text{CR}(\mu-e, \text{N}) = \frac{2\,G_F^2\,\alpha_{w}^2\,m_\mu^5}{(4\pi)^2 \, \Gamma_\text{capt}(Z)} \left|4\,V^{(p)}\left(2 \,\tilde{F}_{u}^{\mu e}+\tilde{F}_{d}^{\mu e}\right) + 4\,V^{(n)}\left(\tilde{F}_u^{\mu e}+2\,\tilde{F}_{d}^{\mu e}\right) + D \,G^{\mu e}_{\gamma} \frac{s^2_w}{2 \sqrt{4 \pi \alpha}}  \right|^2,
\end{equation}
where $\Gamma_\text{capt}(Z)$ denotes the capture rate of a nucleus with atomic number $Z$~\cite{Kitano:2002mt}, $G_F$ is the Fermi constant, $m_\mu$ the muon mass, $\alpha \equiv e^2/(4\pi)$, with $s_w$ corresponding to the sine of the weak mixing angle. The form factors $\tilde{F}_{q}^{\mu e}$ ($q=u,d$) are given by 
\begin{equation}
    \tilde F_q^{\mu e} = Q_q \, s_w^2 F^{\mu e}_\gamma+F^{\mu e}_Z \left(\frac{{I}^3_q}{2}-Q_q\, s_w^2\right) + \frac14 F^{\mu eqq}_\text{box}\,,
    \label{eq:tildeFqmue}
\end{equation}
where $Q_q$ denotes the quark electric charge ($Q_u=2/3$, $Q_d=-1/3$) and ${I}^3_q$ is the weak isospin ($\mathcal{I}^3_u=1/2$, $\mathcal{I}^3_d=-1/2$). The quantities $F^{\mu e}_\gamma$, $F^{\mu e}_Z$ and $F^{\mu eqq}_\text{box}$ correspond to the different form factors of the diagrams, and $G^{\mu e}_\gamma$ corresponds to the dipole term; all expressions are collected in Appendix~\ref{formfactors}. The relevant nuclear information (nuclear form factors and averages over the atomic electric field) is encoded in the form factors $D$, $V^{(p)}$, and $V^{(n)}$. In our analysis, we use the numerical values presented in Ref.~\cite{Kitano:2002mt}.

\subsubsection{\boldmath $\ell_\alpha\to \ell_\beta\ell_\beta\ell_\rho$}
The expression for BR$(\mu \to eee)$~\cite{Ilakovac:1994kj, Alonso:2012ji} we use is 
\begin{align} \label{eq:mueee}
    \text{BR}(\mu \to eee)=& \frac{\alpha^4_w}{24576\pi^3}\frac{m^4_\mu}{M^4_W}\frac{m_\mu}{\Gamma_\mu} \times \Bigg\{ 2 \left|\frac12F^{\mu eee}_{\rm Box}+F^{\mu e}_Z-2s^2_w(F^{\mu e}_Z-F^{\mu e}_\gamma)\right|^2+4 s^4_w \left|F^{\mu e}_Z-F^{\mu e}_\gamma\right|^2 \nonumber \\
    &+ 16 s^2_w \text{Re}\left[	(F^{\mu e}_Z +\frac12F^{\mu eee}_{\rm Box})	G^{\mu e*}_\gamma \right] - 48 s^4_w \text{Re}\left[(F^{\mu e}_Z-F^{\mu e}_\gamma)	G^{\mu
    e*}_\gamma \right]	\nonumber \\ 
    &+32 s^4_w |G^{\mu e}_\gamma|^2\left[\ln \frac{m^2_\mu}{m^2_{e}} -\frac{11}{4}	\right] \Bigg\}, 
\end{align}
which contains the same form factors as those entering in CR($\mu-e$, N), although in different combinations, see Appendix~\ref{formfactors}.

\section{Leptogenesis} \label{lepto} 
In this section, we will analyze the implications of our model for the baryon asymmetry of the Universe and the possibility of generating it via leptogenenesis. In the following, we will only consider its viability in the presence of out-of-equilibrium CP and lepton number violating processes. We will not solve the Boltzmann equations, which deserve a dedicated study beyond the scope of this work. Instead, we will discuss whether or not the regimes explaining mixings and light neutrino masses also fulfill the first necessary conditions for a viable leptogenesis. 

To simplify our analysis, we assume that $M$ is a diagonal matrix and we consider the case where $\left\vert M_{11}\right\vert \ll \left\vert M_{22}\right\vert $. We further assume that the gauge singlet neutral lepton $\Omega_L$ as well as the dark scalar singlet $\varphi_1$ are heavier than the lightest pseudo-Dirac fermions $N_1^\pm=N^\pm$, while for simplicity we work on the basis of a diagonal SM charged lepton mass matrix. In the scenario mentioned above, only the first generation of $N_k^{\pm}$ ($k=1,2$) can contribute to the BAU. Then, the lepton asymmetry parameter, which is induced by the decay process of $N^\pm$, is~\cite{Gu:2010xc, Pilaftsis:1997jf}
\begin{equation}
    \varepsilon_\pm \equiv \dsum\limits_{i=1}^{3}\frac{\left[ \Gamma \left(N_{\pm}\to \nu_{i}h\right) -\Gamma \left(N_\pm\to \overline{\nu }_{i}h\right) \right] }{\left[ \Gamma \left(N_\pm\to \nu_{i}h\right) +\Gamma \left(N_\pm\to \overline{\nu }_{i}h\right) \right] }\simeq \frac{\func{Im}\left\{ \left(\left[ \left(y_{N_{+}}\right)^\dagger \left(y_{N_{-}}\right) \right]^2\right)_{11}\right\} }{8\pi A_\pm}\frac{r}{r^2+\frac{\Gamma_\pm^2}{m_{N_\pm}^2}} \,,
\end{equation}
with
\begin{align}
    r = \frac{m_{N_{+}}^2-m_{N_{-}}^2}{m_{N_{+}} m_{N_{-}}},\hspace{4.7cm} &A_\pm=\left[ \left(y_{N_\pm}\right)^\dagger y_{N_\pm}\right]_{11}, \\
    y_{N_\pm} = y_\nu \left(1\mp S\right) =y_\nu \left(1\pm \frac14 M^{-1}\mu \right) \hspace{1.4cm} &\Gamma_\pm=\frac{A_\pm m_{N_\pm}}{8\pi}\,.
\end{align}

Neglecting the interference terms involving the two different sterile
neutrinos $N^\pm$, the washout parameter $K_{N^{+}}+K_{N^{-}}$ is huge, as mentioned in Ref.~\cite{Dolan:2018qpy}. However, small mass splitting between pseudo-Dirac neutrinos leads to destructive interference in the scattering process~\cite{Blanchet:2009kk}. The washout parameter including the interference term has the form
\begin{equation}
    K^\text{eff} \simeq K_{N^{+}}\, \delta_{+}^2 + K_{N^{-}}\, \delta_{-}^2\,,
\end{equation}
where
\begin{equation}
    \delta_\pm=\frac{m_{N^{+}}-m_{N^{-}}}{\Gamma_{N^\pm}}\,,\hspace{0.7cm} \hspace{0.7cm}K_{N^\pm}=\frac{\Gamma_\pm}{H(m_{N_\pm})}\,,
\end{equation}%
and $H$ corresponds to the Hubble expansion rate of the universe.
In the case of a standard cosmological scenario where the total energy density is dominated by SM radiation,
\begin{equation} \label{eq: Hubble}
    H(T) = \frac{\pi}{3}\, \sqrt{\frac{g_\star}{10}}\, \frac{T^2}{M_P}\,,
\end{equation}
where $g_\star$ corresponds to the number of relativistic degrees of freedom in the SM bath, and $M_P \simeq 2.4 \times 10^{18}$~GeV is the reduced Planck mass.

In the weak and strong washout regimes, the BAU is related to the lepton asymmetry~\cite{Pilaftsis:1997jf} as follows 
\begin{align}
    Y_{\Delta B} &\equiv \frac{n_{B}-\overline{n}_{B}}{s}=-\frac{28}{79}\frac{\epsilon_{+}+\epsilon_{-}}{g^*},\hspace*{2.2cm}\text{for}\hspace*{0.5cm}K^\text{eff}\ll 1, \\
    Y_{\Delta B} &\equiv \frac{n_{B}-\overline{n}_{B}}{s}=-\frac{28}{79}\frac{0.3\left(\epsilon_{+}+\epsilon_{-}\right) }{g^*K^\text{eff}\left(\ln K^\text{eff}\right)^{0.6}}\,,\hspace*{0.5cm}\text{for}\hspace*{0.5cm}K^\text{eff}\gg 1.
\end{align}
In the analysis, we constrain the model in order to successfully reproduce the experimental value for the baryon asymmetry parameter~\cite{Planck:2018vyg}
\begin{equation}
    Y_{\Delta B} = \left(0.87\pm 0.01\right) \times 10^{-10}.
\end{equation}

\section{Dark matter relic density} \label{Sec:DM}
The residual symmetry protects the lightest odd state of the dark sector, rendering it a viable candidate for DM. Depending on the mass hierarchy, we could have scalar or fermionic DM candidates. To simplify our analysis, we assume without loss of generality that $\varphi_{2DM}$ is sufficiently heavy to allow its decay, then implying that the lightest among the $\varphi_{1R}$, $\varphi_{1I}$, that is, $\varphi_{1DM}$ is the scalar DM candidate.  In addition to that, in this scenario, the DM candidate scalar $\varphi_{1DM}$ will annihilate into a couple of SM particles and pairs of BSM scalars $\sigma_R\sigma_R$, $\sigma_I\sigma_I$ through the Higgs portal scalar interactions $\lambda_{6}(\phi^\dagger \phi )(\varphi_{1}^*\varphi_{1})$ and $\lambda_{8}(\sigma^*\sigma )(\varphi_{1}^*\varphi_{1})$, respectively.  Furthermore, the scalar DM candidate $\varphi_{1DM}$ can also annihilate into neutrino pairs $\widetilde{\nu}_{i}\widetilde{\nu}_{j}$ (active-active), $\widetilde{\nu}_{i}N_{j}^{(\pm)}$ (active-heavy),  $N_{i}^{(\pm)}N_{j}^{(\pm)}$ (heavy-heavy) ($i,j=1,2,3$) and $\widetilde{\Psi}^{(\pm)}_k\widetilde{\Psi}^{(\pm)}_r$ (where $k,r=1,2$ and $\widetilde{\Psi}^{(\pm)}_k$ are the physical states arising from $m_{\Psi}$ appearing in the second term of Eq.~\eqref{neutrinomassterms}) via the $t$ channel exchange of the heavy neutral leptons $\widetilde{\Psi}^{(\pm)}_k$ ($k=1,2$) and neutrinos $\widetilde{\nu}_{i}$ and $N_{i}^{(\pm)}$ ($i=1,2,3$), respectively. Moreover, $\varphi_{1DM}$ can also annihilate in neutrino pairs $\widetilde{\Psi}^{(\pm)}_k\widetilde{\Psi}^{(\pm)}_r$ and $\Omega_k\Omega_r$ via the $t$ channel exchange of $\Omega_k$ and $\widetilde{\Psi}^{(\pm)}_k$ ($k,r=1,2$), respectively. These annihilation channels will contribute to the DM relic density, which can be successfully accommodated for appropriate values of the scalar DM mass and couplings of the aforementioned scalar portal interactions.

Regarding the fermionic DM possibility, which we focus on from now on, $\widetilde{\Psi}^{(\pm)}_2$. Here in this work, to simplify our analysis, we assume that $\Omega_k$ ($k=1,2$) is sufficiently heavy to allow its decay into the $\varphi_2\Psi_{nL}$ final state, then implying that the lightest stable candidates are $\widetilde{\Psi}^{(\pm)}_1$ having a common mass $m_{\widetilde{\Psi}_{1}}$.  This case is particularly interesting because the Yukawa coupling that mediates DM annihilation also participates in the calculation of the $\mu$ parameter, allowing us to correlate the cFLV, DM, and neutrino masses.  In this case, for DM masses $m_{\widetilde{\Psi}_1}$ in the GeV-to-TeV ballpark and Yukawa couplings at the electroweak scale, DM could have been generated in the early Universe via the WIMP mechanism.
The evolution of the DM number density $n$ can be tracked with the help of the Boltzmann equation
\begin{equation}
    \frac{dn}{dt} + 3\, H\, n = -\langle\sigma v\rangle \left(n^2 - n_\text{eq}^2\right),
\end{equation}
where $\langle\sigma v\rangle$ is the total thermally-averaged annihilation cross section of a couple of DM particles into lighter states, $n_\text{eq}(T)$ corresponds to the DM number density in equilibrium at a temperature $T$, given by
\begin{equation}
    n_\text{eq}(T) \simeq \left(\frac{m_{\widetilde{\Psi}_{1}}\, T}{2 \pi}\right)^{3/2} e^{-\frac{m_{\widetilde{\Psi}_{1}}}{T}}
\end{equation}
in the nonrelativistic limit, and $H$ corresponds to the Hubble expansion rate of the Universe assumed to be dominated by SM radiation.\footnote{DM freeze out in nonstandard cosmological scenarios or during a low-temperature reheating era is also possible, but will not be discussed here~\cite{Allahverdi:2020bys}.}

In the early Universe, pairs of non-relativistic DM $\widetilde{\Psi}^{(\pm)}_1$ can annihilate mainly into right-handed neutrinos $N_{nR}$ through the $t$-channel exchange of $\varphi_1$. Later, $N_{nR}$ decays into Higgs and lepton doublets.
The corresponding squared amplitude for the annihilation is given by
\begin{equation}
        |\mathcal{M}|^2 =\frac{1}{2} \left|(y_N)_{n1}\right|^4 \left(\frac{t - \left(m_{\widetilde{\Psi}_{1}} + m_{N_{nR}}\right)^2}{t - m_{\varphi_1}^2}\right)^2,
\end{equation}
and hence the thermally-averaged annihilation cross section $\langle\sigma v\rangle$ is
\begin{equation}
    \langle\sigma v\rangle = \frac{T}{64 \pi^4\, n_\text{eq}^2(T)} \int_{4 m_{\widetilde{\Psi}_{1}}^2}^\infty ds\, \sqrt{s}\, \sigma_R(s)\, K_1\left(\frac{\sqrt{s}}{T}\right),
\end{equation}
where $K_1$ is the modified Bessel function and $\sigma_R(s)$ is the reduced cross section
\begin{equation}
    \sigma_R(s) \equiv \frac{1}{8\pi\, s} \int |\mathcal{M}|^2\, dt\,.
\end{equation}
It is interesting to note that $\widetilde{\Psi}^{(\pm)}_1$ can co-annihilate with $\widetilde{\Psi}^{(\pm)}_2$ if their mass difference is typically smaller than $\sim 20\%$~\cite{Griest:1990kh}. In that case, an additional Boltzmann equation for the other $\widetilde{\Psi}^{(\pm)}_2$ must be taken into account. Here, however, we assume a larger splitting so that coannihilation processes can safely be ignored.

To match the entire observed DM relic density, it is required that
\begin{equation}
    m_{\widetilde{\Psi}_{1}}\, \frac{n_0}{s_0} = \frac{\Omega h^2\, \rho_c}{s_0\, h^2} \simeq 4.3 \times 10^{-10}~\text{GeV},
\end{equation}
where $n_0/s_0$ is the asymptotic value of the DM yield at low temperatures, $s_0 \simeq 2.69 \times 10^3$~cm$^{-3}$ is the present entropy density~\cite{ParticleDataGroup:2022pth}, $\rho_c \simeq 1.05 \times 10^{-5}~h^2$~GeV/cm$^3$ is the critical energy density of the Universe, and $\Omega h^2 \simeq 0.12$ is the observed DM relic abundance~\cite{Planck:2018vyg}.

\section{\boldmath Interplay between DM, cLFV, LNV, $0\nu\beta\beta$ and leptogenesis} \label{sec:interplay}
In this section, we discuss the implications of the considered model for DM, cLFV, LNV, $0\nu\beta\beta$, and leptogenesis. To this end, we performed a random scan in the ranges $m_{\Psi_{1R}} \in [50,\, 500]$~GeV, $m_{\varphi_1} \in[600,\, 2000]$~GeV and $m_{N_{1R}} \in [10~\text{GeV},\, m_{\Psi_{1R}}/1.1]$ and computed the Yukawa coupling $Y _{N _{11}}$ required to fit the entire observed abundance of DM. Note that from now on we call $m_{\widetilde{\Psi}_{k}}$ as $m_{\Psi_{kR}}$ ($k=1,2$).  The other parameters, relevant for computing the neutrino masses and other observables, are randomly varied in the ranges $m _{\Psi _{2R}} \in [m _{\Psi _{1R}}, 300~\text{GeV}]$, $m _{\varphi _2} \in [m _{\varphi _1}, 5~\text{TeV}]$, $m_{N_2}, m_\Omega \in [700~\text{GeV}, 5~\text{TeV}]$, $y _{N _{nk}}$, $y _{\Omega _{nk}} \in [-2,\, 2]$ and $\lambda _{10} \in [0.1,\, 2]$. Throughout our analysis, we impose agreement with the light-neutrino data. The consistency with the measured values of neutrino mass squared splittings, leptonic mixing angles, and leptonic Dirac CP-violating phase arising from neutrino oscillation experiments is guaranteed by the use of the Casas-Ibarra parameterization of the ISS mechanism, which provides the values of the entries of the Dirac neutrino submatrix, given the Majorana submatrices $M$ and $\mu$, as indicated in Eq.~\eqref{mnuD}.

\begin{figure}
    \begin{center}
    \includegraphics[width=0.49\textwidth]{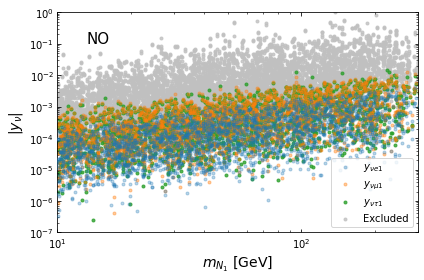}
    \includegraphics[width=0.49\textwidth]{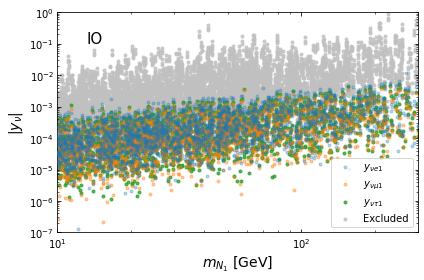}
    \end{center}
    \caption{Allowed values in the $|y_{\nu}|$-$m_{N_1}$ plane consistent with the DM relic abundance and neutrino oscillation data assuming normal (inverted) ordering in the left (right) panel. The small $m_{N_1}$ leads to Yukawa couplings smaller than $10^{-2}$, to generate the correct active neutrino masses. Higher Yukawa couplings, although still allowed by oscillation data, are excluded by cLFV constraints (gray points).}
    \label{yvsmN}
\end{figure}
We show in Fig.~\ref{yvsmN} the allowed parameter space in the $|y_{\nu}|$-$m_{N_1}$ plane consistent with the constraints imposed by the measured value of the DM relic abundance and by the experimental neutrino oscillation data. We consider both normal and inverted neutrino mass hierarchies, corresponding to the left and right panels of Fig.~\ref{yvsmN}, respectively. The consistency with the experimental values of the neutrino mass squared differences and with the constraints arising from the charged-lepton violation requires Yukawa couplings $y _{\nu _{ik}}$ associated with the Dirac submatrix smaller than $10^{-2}$. On the other hand, the gray points in Fig.~\ref{yvsmN} correspond to large mixing angles between active and heavy neutrinos that give rise to unacceptably large rates for the lepton flavor-violating process, higher than their corresponding upper experimental limits, and thus they are excluded by cLFV constraints.

\begin{figure}[t]
    \begin{center}
    \includegraphics[width=0.49\textwidth]{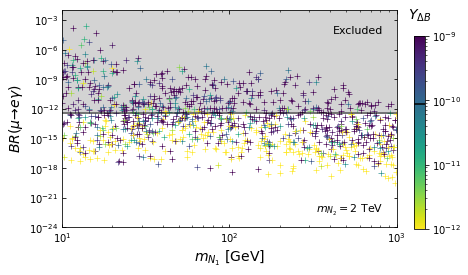} 
    \includegraphics[width=0.49\textwidth]{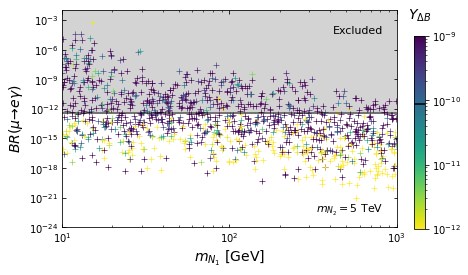}
    \end{center}
    \caption{$BR(\mu \to e \gamma)$ as a function of $m _{N _1}$ for fixed values of $m _{N _2}$ at 2 TeV and 5 TeV. The points reproduce neutrino oscillation data (using the $\mu$ parameterization), and are colored according to the size of the baryon asymmetry parameter $Y_{\Delta B}$. Points that have $Y_{\Delta B}$ of the order of the correct experimental value typically pass the constraint of $\mu \to e \gamma$, shown by the horizontal black line.}
    \label{Mrmutoegamma}
\end{figure}
Figure~\ref{Mrmutoegamma} shows the values obtained for the branching ratio for the decay of $\mu\to e\gamma$ for random values of the entries of the Dirac submatrix $m_{\nu D}$ and different masses of the lightest pseudo-Dirac neutral lepton $N_1$ with $m_{N_2}$ fixed to be equal to 2~TeV and 5~TeV, for the left and right panels, respectively. Consistency with the neutrino oscillation data in Fig.~\ref{Mrmutoegamma} is ensured by using the $\mu$ parameterization, which implies that to successfully reproduce the experimental values of the neutrino mass squared splittings, leptonic mixing parameters, and the leptonic Dirac CP phase, the Majorana submatrix $\mu$, in the basis of diagonal SM charged lepton mass matrix, should have the form~\cite{MarcanoImaz:2017xjc, Hernandez:2021uxx}
\begin{equation}
    \mu =M^{T}m_{\nu D}^{-1}\widetilde{\mathbf{M}}_\nu \left( m_{\nu D}^{T}\right) ^{-1}M=M^{T}m_{\nu D}^{-1}U_\text{PMNS}\left( \widetilde{\mathbf{M}}_\nu \right)_\text{diag} U_\text{PMNS}^{T}\left(m_{\nu D}^{T}\right) ^{-1}M,
\end{equation}
where $\left( \widetilde{\mathbf{M}}_\nu \right)_\text{diag}$ defined in Eq.~\eqref{eq:Mdiag}, and $U_\text{PMNS}$ the PMNS leptonic mixing matrix (without unitarity violations). It proves convenient to employ the $\mu$ parameterization in this case to avoid the apparent non-decoupling behavior with the mass of the heavy neutrinos when using the Casas-Ibarra parameterization, which would lead to a constant value of the cLFV rates for large $m _{N _1}$ \cite{MarcanoImaz:2017xjc}. The intensity of the colors in Fig.~\ref{Mrmutoegamma} corresponds to different values of the baryon asymmetry parameter. As shown in Fig.~\ref{Mrmutoegamma}, there are several points corresponding to the decay rate $\mu\to e\gamma$ lower than its upper experimental limit of $4.2\times 10^{-13}$ and that yields values for the baryon asymmetry parameter consistent with its experimental value.

\begin{figure}[t!]
    \begin{center}
    \includegraphics[width=0.49\textwidth]{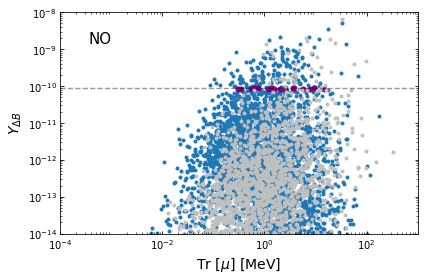}
    \includegraphics[width=0.49\textwidth]{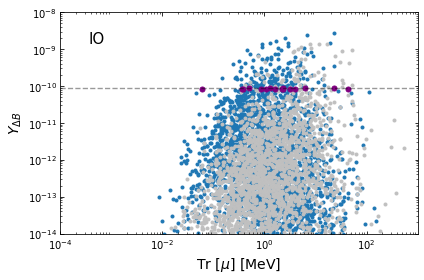}
    \end{center}
    \caption{Baryon asymmetry parameter $Y_{\Delta B}$ as a function of the trace of the Majorana submatrix $\mu$. All points comply with the DM relic abundance and neutrino oscillation data assuming normal (inverted) ordering in the left (right) panel. The purple points have the correct $Y_{\Delta B}$ within $3 \sigma$, while the gray points are excluded by cLFV. The values of $\mu$ in the keV-MeV range are favored by leptogenesis.}
    \label{YBvsTrmu}
\end{figure}
Figure~\ref{YBvsTrmu} shows the baryon asymmetry parameter as a function of the trace of the Majorana submatrix $\mu$ for scenarios of normal (left panel) and inverted (right panel) neutrino mass hierarchies. The consistency with the neutrino oscillation data is ensured in this case by the Casas-Ibarra parameterization. The gray points in Fig.~\ref{YBvsTrmu} are excluded by the constraints arising from cLFV. The purple points correspond to the values of the baryon asymmetry parameter $Y_{\Delta B}$ within the experimentally allowed range at $3 \sigma$~CL. As shown in Fig.~\ref{YBvsTrmu}, this model  is consistent with the experimental range of the baryon asymmetry parameter, provided that the entries of the Majorana submatrix $\mu$ are in the keV to MeV range.

\begin{figure}[t!]
    \begin{center}
    \includegraphics[width=0.49\textwidth]{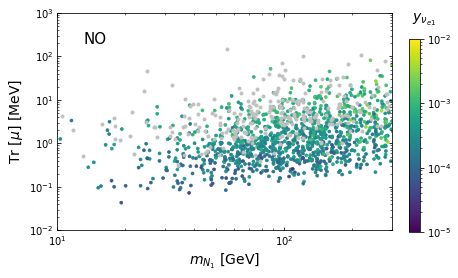}
    \includegraphics[width=0.49\textwidth]{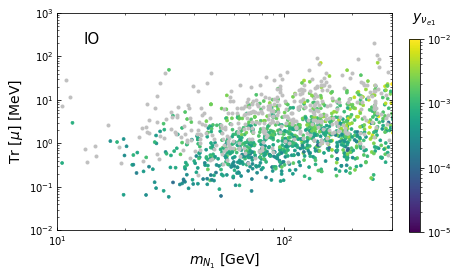}
    \end{center}
    \caption{Correlation between the trace of the Majorana submatrix $\mu$ and $m_{N_1}$. Points comply with the BAU, DM relic abundance and neutrino oscillation data assuming normal (inverted) ordering in the left (right) panel. The points are colored according to the size of $y _{\nu _{e 1}}$. Gray points are excluded by cLFV constraints.}
    \label{TrmuvsmN1}
\end{figure}

Having studied the parameter space favored by DM and leptogenesis separately, we next combine this information to show that both cosmological and electroweak precision data can be successfully accommodated at the same time. Again, the experimental neutrino oscillation data are also guaranteed by the Casas-Ibarra parameterization. Figure~\ref{TrmuvsmN1} shows solutions in the plane Tr$[\mu]$-$m _{N _1}$ that reproduce the DM relic abundance and the BAU simultaneously. The colors correspond to the values of the $y _{\nu _{e1}}$ Dirac Yukawa coupling. The gray points in Fig.~\ref{TrmuvsmN1} correspond to rates for cLFV processes larger than their experimental upper limits and are therefore excluded.

\begin{figure}[t!]
    \begin{center}
    \includegraphics[width=0.49\textwidth]{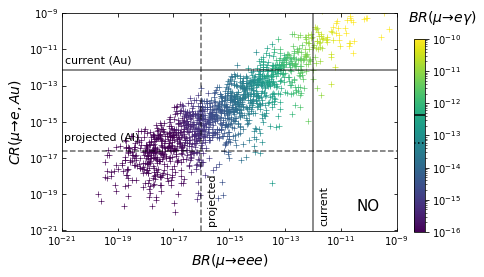}
    \includegraphics[width=0.49\textwidth]{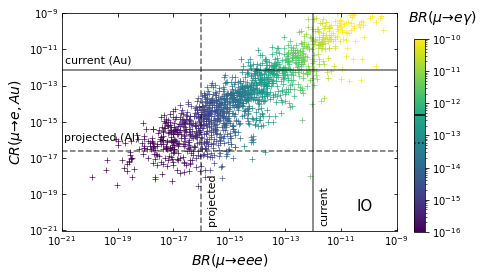}
    \end{center}
    \caption{Correlation among cLFV observables. Points comply with the BAU, DM relic abundance and neutrino oscillation data assuming normal (inverted) ordering in the left (right) panel. Current bounds are shown as full lines, while projections are shown as dashed lines.}
    \label{cLFV}
\end{figure}
The correlation among the cLFV observables for the scenarios of normal and inverted neutrino mass hierarchies is shown in the left and right panels of Fig.~\ref{cLFV}, respectively. As indicated in Fig.~\ref{cLFV}, an increase of the $\mu\to eee$ decay rate produces larger values for the $\mu\to e$ conversion rate as well as for the $\mu\to e\gamma$ branching ratio. Current bounds are shown with full lines, while projections are shown with dashed lines. All the points shown in Fig.~\ref{cLFV} agree with the experimental values of the DM relic abundance, BAU, and neutrino oscillation. Figure~\ref{cLFV} shows a nice interplay between cosmological and particle physics observables. Although it shows that the cLFV constraints already exclude many otherwise viable points to account for DM and leptogenesis, it also shows that the upcoming cLFV experiments will be able to probe a large portion of the remaining ones.

It is worth mentioning that despite the large number of similarities in the different correlation plots among several observables, in the plot of Fig.~\ref{TrmuvsmN1}, which displays the correlation between the trace of the Majorana submatrix $\mu$ and the mass $m_{N_1}$ of the lightest pseudo-Dirac lepton, in the region of large $m_{N_1}$ values, there are more gray points (points excluded by the cLFV constraints) for the case of inverted hierarchy than for the normal neutrino mass spectrum.

\section{Other three-loop inverse seesaw models} \label{sec:Examples}
For completeness, here we present two more versions of the model setup leading to three-loop ISS neutrino mass generation. We do not explore the phenomenological and cosmological implications of these models here and leave this task to future work.

\subsection{Model 2} \label{Sec:model1}
This second example is based on the same gauge group of Eq.~\eqref{eq:Symmetry-1}, but with a different spontaneous symmetry breaking scheme 
\begin{align}
    \label{eq:Symm-2}
    &SU(3)_{C}\otimes SU(2)_L\otimes U(1)_Y\otimes U(1)' \otimes \mathbb{Z}_2 \\
    &\hspace{22mm}\Downarrow v_\sigma  \notag \\
    &SU(3)_{C}\otimes SU(2)_L\otimes U(1)_Y \otimes \mathbb{Z}_2  \notag\\
    &\hspace{22mm}\Downarrow v_{\phi}  \notag \\
    &SU(3)_{C}\otimes U(1)_\text{em}\otimes \mathbb{Z}_2\,. \nonumber
\end{align}
The global $U(1)'$ symmetry is spontaneously broken at the TeV scale by the VEV of $\langle\sigma\rangle$. 
Other new scalars with non-trivial charges $\mathbb{Z}_2$ do not acquire VEVs to maintain this unbroken symmetry. The charges for the fields under the symmetry of Eq.~\eqref{eq:Symm-2} are shown in Table~\ref{model1}.
The neutrino Yukawa interations invariant with respect to this group are 
\begin{align}
    -\mathcal{L}_Y^{(f)} = &\sum_{i=1}^3  \sum_{k=1}^2  \left(y_{\nu}\right)_{ik}\, \overline{l}_{iL}\, \widetilde{\phi}\, \nu_{kR} + \sum_{n=1}^2  \sum_{k=1}^2  M_{nk}\, \overline{\nu}_{nR}\, N_{kR}^C \notag \\
    &\quad + \sum_{n=1}^2  \sum_{k=1}^2  \left(y_{N}\right)_{nk}\, \overline{N}_{nR}\, \varphi\, \Psi_{kR}^C + \sum_{n=1}^2  \sum_{k=1}^2 \left(m_{\Psi}\right)_{nk}\, \overline{\Psi}_{nR}\, \Psi_{kR}^C + \text{H.c.}
\end{align}

\begin{figure}[t]
    \begin{center}
        \includegraphics[width=0.65\textwidth]{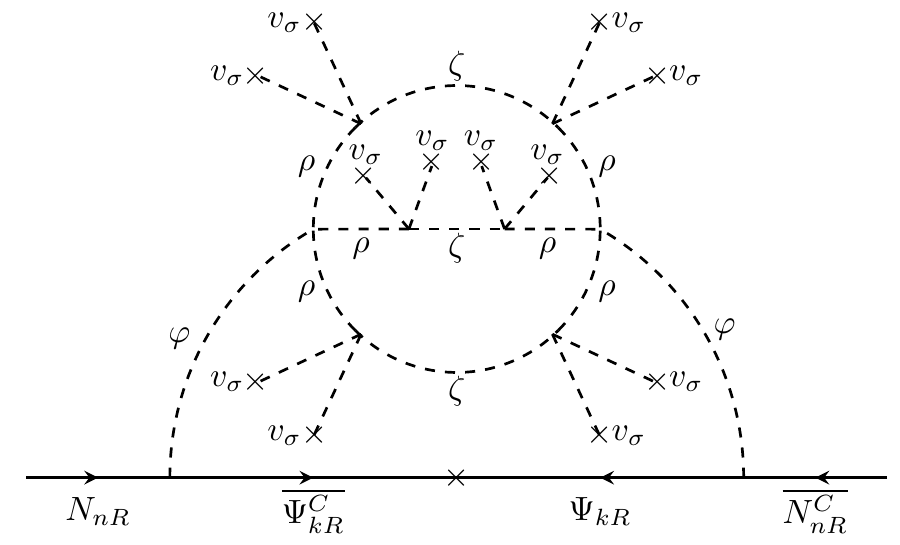}
    \end{center}
    \caption{Three-loop diagram for the LNV mass term in model 2, with $n, k = 1, 2$.}
    \label{Neutrinodiagram2}
\end{figure}

Note that the mass splitting between the real and imaginary parts of the scalar singlet $\varphi$ is generated only at the two-loop level and therefore is small. With this at hand and assuming that the neutral leptons $\Psi_{kR}$ are heavy enough, we can dynamically generate the small Majorana mass terms $\left(\mu \right)_{nk}\overline{N}_{nR}N_{kR}^C$ at the three-loop level, as shown in Fig.~\ref{Neutrinodiagram2}. The preserved $\mathbb{Z}_2$ discrete symmetry allows for a stable scalar or fermionic DM candidate corresponding to the lightest electrically neutral particle odd under $\mathbb{Z}_2$. The fermionic DM candidate can be the lightest of the two $\Psi_{kR}$ states.
{\Large \begin{table}[t]\centering
    \renewcommand{\arraystretch}{1.3} 
    \begin{tabular}{|c|c|c|c|c|c|c|c|c|c|c|}
\hline
Field & $l_{iL}$ & $l_{iR}$ & $\nu _{kR}$ & $N_{kR}$ & $\Psi _{kR}$ & $\phi $
& $\varphi $ & $\rho $ & $\zeta $ & $\sigma $ \\ \hline\hline
$SU(3)_{C}$ & $\mathbf{1}$ & $\mathbf{1}$ & $\mathbf{1}$ & $\mathbf{1}$ & $%
\mathbf{1}$ & $\mathbf{1}$ & $\mathbf{1}$ & $\mathbf{1}$ & $\mathbf{1}$ & $%
\mathbf{1}$ \\ \hline
$SU(2)_{L}$ & $\mathbf{2}$ & $\mathbf{1}$ & $\mathbf{1}$ & $\mathbf{1}$ & $%
\mathbf{1}$ & $\mathbf{2}$ & $\mathbf{1}$ & $\mathbf{1}$ & $\mathbf{1}$ & $%
\mathbf{1}$ \\ \hline
$U(1)_{Y}$ & $-\frac{1}{2}$ & $-1$ & $0$ & $0$ & $0$ & $\frac{1}{2}$ & $0$ & 
$0$ & $0$ & $0$ \\ \hline
$U(1)^{\prime }$ & $0$ & $3$ & $-3$ & $3$ & $0$ & $-3$ & $3$ & $-1$ & $0$ & $%
\frac{1}{2}$ \\ \hline
$\mathbb{Z}_{2}$ & $0$ & $0$ & $0$ & $0$ & $1$ & $0$ & $1$ & $1$ & $1$ & $0$
\\ \hline
\end{tabular}
    \caption{Model 2. Charge assignments under $SU(3)_C\otimes SU(2)_L\otimes U(1)_Y\otimes U(1)'\otimes \mathbb{Z}_2$ symmetry. Here $i = 1$, 2, 3 and $k = 1$, 2.}
    \label{model2}
\end{table}}

\subsection{Model 3} \label{Sec:model2}
Model 3 is similar to Model 2, with the important difference that an additional $U(1)_{B-L}$ gauge symmetry is introduced, together with $U(1)'\otimes \mathbb{Z}_2$ symmetry. This model, compared to Model 1,  has one extra scalar and one RH Majorana neutrino, which means that in total it contains five electrically neutral scalar singlets $\varphi$, $\rho$, $\zeta$, $\sigma$, $\chi$, and six RH neutrinos $\nu_{kR}$, $N_{kR}$, $\Psi_{iR}$ (with $k = 1$, 2 and $i = 1$, 2, 3). The increased number of fermions comes from the requirement of cancelation of chiral anomalies. The quark and lepton fields and their assignments are shown in Table~\ref{model3}. Note that, in contrast to Models 1 and 2, the quark and lepton sectors are correlated here by the condition of anomaly cancelation. The corresponding neutrino Yukawa interactions are given by the Lagrangian density
\begin{align}
    -\mathcal{L}_Y^{\left(\nu\right)} = &\sum_{i=1}^3  \sum_{k=1}^2 \left(y_{\nu}\right)_{ik}\, \overline{l}_{iL}\, \widetilde{\phi}\, \nu_{kR} + \sum_{n=1}^2 \sum_{k=1}^2 M_{nk}\, \overline{\nu }_{nR}\, N_{kR}^C \notag \\
    &+ \sum_{n=1}^2 \sum_{j=1}^3  \left(y_{N}\right)_{nj}\, \overline{N}_{nR}\, \varphi\, \Psi_{jR}^C + \sum_{i=1}^3  \sum_{j=1}^3  \left(y_{\Psi}\right)_{ij}\, \overline{\Psi}_{iR}\, \chi\, \Psi_{jR}^C + \text{H.c.}
\end{align}
{\Large{\begin{table}[t!]\centering
    \renewcommand{\arraystretch}{1.3} 
    \begin{tabular}{|c||c|c|c|c|c|c|c|c|c|c|c|c|c|c|}
        \hline
        Field & $q_{iL}$ & $u_{iR}$ & $d_{iR}$ & $l_{iL}$ & $l_{iR}$ & $\nu_{kR}$ & $N_{kR}$ & $\Psi_{iR}$ & $\phi $ & $\varphi $ & $\rho $ & $\zeta $ & $\sigma $ & $\chi$ \\ \hline\hline
        $SU(3)_C$ & $\mathbf{3}$ & $\mathbf{3}$ & $\mathbf{3}$ & $\mathbf{1}$ & $\mathbf{1}$ & $\mathbf{1}$ & $\mathbf{1}$ & $\mathbf{1}$ & $\mathbf{1}$ & $\mathbf{1}$ & $\mathbf{1}$ & $\mathbf{1}$ & $\mathbf{1}$ & $\mathbf{1}$ \\ \hline
        $SU(2)_L$ & $\mathbf{2}$ & $\mathbf{1}$ & $\mathbf{1}$ & $\mathbf{2}$ & $\mathbf{1}$ & $\mathbf{1}$ & $\mathbf{1}$ & $\mathbf{1}$ & $\mathbf{2}$ & $\mathbf{1}$ & $\mathbf{1}$ & $\mathbf{1}$ & $\mathbf{1}$ & $\mathbf{1}$ \\ \hline
        $U(1)_Y$ & $\frac16$ & $\frac23$ & $-\frac13$ & $-\frac12 $ & $-1$ & $0$ & $0$ & $0$ & $\frac12 $ & $0$ & $0$ & $0$ & $0$ & $0$ \\ \hline
        $U(1)_{B-L}$ & $\frac13$ & $\frac13$ & $\frac13$ & $-1$ & $-1$ & $-1$ & $1$ & $-1$ & $0$ & $0$ & $0$ & $0$ & $0$ & $2$ \\ \hline
        $U(1)'$ & $0$ & $-4$ & $4$ & $0$ & $4$ & $-4$ & $4$ & $-1$ & $-4$ & $3$ & $-1$ & $0$ & $\frac12 $ & $2$ \\ \hline
        $\mathbb{Z}_2$ & $0$& $0$& $0$& $0$& $0$& $0$& $0$& $1$& $0$& $1$& $1$& $1$& $0$& $0$\\ \hline
    \end{tabular}
    \caption{Model 3. Charge assignments under the $SU(3)_C \otimes SU(2)_L \otimes U(1)_Y \otimes U(1)_{B-L} \otimes U(1)' \otimes \mathbb{Z}_2$ symmetry. Here $i=1, 2, 3$ and $k = 1$, 2.}
    \label{model3}
\end{table}}}

In our model the scalar fields $\sigma, \chi, \phi$ develop non-zero VEVs $\langle\sigma\rangle, \langle\chi\rangle, \langle\phi\rangle$ breaking the model symmetry in two ways: for $v_\chi < v_\sigma$ 
\begin{align}
    & SU(3)_C \otimes SU(2)_L \otimes U(1)_Y \otimes U(1)_{B-L} \otimes U(1)' \otimes \mathbb{Z}_2 \notag \\
    &\hspace{35mm}\Downarrow v_\sigma  \notag \\
    &SU(3)_C \otimes SU(2)_L \otimes U(1)_Y \otimes U(1)_{B-L}  \otimes \mathbb{Z}_2 \notag \\
    &\hspace{35mm}\Downarrow v_{\chi}  \notag \\
    &SU(3)_{C}\otimes SU(2)_L\otimes  \mathbb{Z}_6 \otimes \mathbb{Z}_2 \notag\\
    \label{eq:SSB-1}
    &\hspace{35mm}\Downarrow v_{\phi}  \notag \\
    &SU(3)_{C}\otimes U(1)_\text{em}\otimes   \mathbb{Z}_6 \otimes \mathbb{Z}_2\,,
\end{align}
and for $v_\chi > v_\sigma$
\begin{align}
    & SU(3)_C \otimes SU(2)_L \otimes U(1)_Y \otimes U(1)_{B-L} \otimes U(1)' \otimes \mathbb{Z}_2 \notag \\
    &\hspace{35mm}\Downarrow v_{\chi}  \notag \\
    &SU(3)_{C}\otimes SU(2)_L\otimes  \mathbb{Z}_6 \otimes \mathbb{Z}'_6 \otimes \mathbb{Z}_2 \notag\\
    &\hspace{35mm}\Downarrow v_\sigma  \notag \\
    &SU(3)_{C}\otimes SU(2)_L\otimes  \mathbb{Z}_6 \otimes \mathbb{Z}_2 \notag\\
    \label{eq:SSB-2}
    &\hspace{35mm}\Downarrow v_{\phi}  \notag \\
    &SU(3)_{C}\otimes U(1)_\text{em}\otimes \mathbb{Z}_6 \otimes \mathbb{Z}_2\,.
\end{align}
In both cases, the residual $\mathbb{Z}_6$ originates from the spontaneously broken $U(1)_{B-L}$ and survives after the electroweak symmetry breaking. In the second case, $v_\chi > v_\sigma$, in the first step of symmetry breaking by $v_{\chi} = \langle\chi\rangle$ there appears another residual symmetry $\mathbb{Z}'_6 \subset U(1)'$, which is, however, completely broken in the second step by $v_\sigma$. 

Note that the $\mathbb{Z}_2$ parity is preserved at low energies, offering stable scalar or fermionic DM candidates. In this model, the scalar fields $\varphi$, $\rho$, and $\zeta$ do not acquire VEVs and as they have non-trivial $\mathbb{Z}_2$ charges the lightest of their real and/or imaginary parts can be a viable DM candidate.

\begin{figure}[t]
    \begin{center}
        \includegraphics[width=0.65\textwidth]{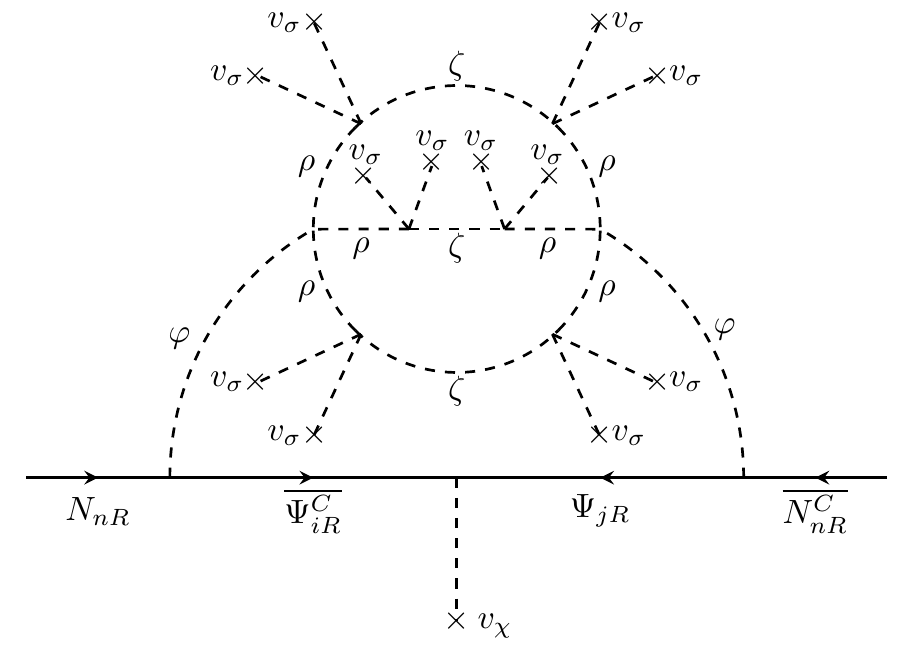}
    \end{center}
    \caption{Three-loop diagram for the LNV mass term in model 3, with $n = 1, 2$ and $i, j = 1, 2, 3$.}
    \label{Neutrinodiagram3}
\end{figure}

In both scenarios (\ref{eq:SSB-1}) and (\ref{eq:SSB-2}), the LNV required for the generation of Majorana neutrinos masses arises due to the spontaneous breaking of $U(1)_{B-L}$ by $\langle\chi\rangle\neq 0$. Thus, in this model, there is an SM singlet majoron $\sim Im$ $\chi$. The LNV comes into play only through the Majorana masses at tree level $\sim y \langle\chi\rangle$ of the seesaw mediators $\Psi_{iR}$, charged under $U(1)_{B-L}$. The LNV is then transferred to active neutrinos at the loop level. Similarly to Model 1, the $U(1)' \otimes \mathbb{Z}_2$ symmetry forbids active neutrino masses at one- and two-loop levels but allows them via ISS with the Majorana mass term $\left(\mu \right)_{nk} \overline{N}_{nR} N_{kR}^C$ dynamically generated at the three-loop level, as shown in Fig.~\ref{Neutrinodiagram3}. 

\section{Conclusions} \label{Sec:conclusions}
We have constructed three models where the tiny masses of the active neutrino are radiatively generated from an inverse seesaw mechanism at three-loop level, thanks to preserved discrete symmetries. In these models, the Majorana mass submatrix corresponding to the lepton number violating (LNV) $\mu$ parameter is radiatively generated via the three-loop level exchange of neutral leptons and gauge singlet scalars. Such a three-loop suppression for the LNV $\mu$ parameter allows values for the entries of the Majorana mass submatrix in the keV to MeV range. Focusing on one of these models, where the SM particle content is extended by the inclusion of several neutral leptons and gauge singlet scalar fields and the SM gauge symmetry is extended with the global symmetry $U(1)' \otimes \mathbb{Z}_2$, we analyzed in detail the scalar sector and explored the implications of the model in neutrino masses and in the lepton sector phenomenology for both scenarios of normal and inverted neutrino mass hierarchies. In such a model, the $\mathbb{Z}_2$ symmetry is preserved, and the $U(1)'$ symmetry is spontaneously broken down to a residual conserved $\mathbb{Z}_4$ symmetry. These discrete symmetries are crucial for ensuring the radiative nature of the inverse seesaw mechanism as well as the stability of fermionic and scalar dark matter candidates, particles that mediate the three-loop level radiative seesaw mechanism that yields the LNV $\mu$ parameter.  We found that the considered model successfully complies with the constraints imposed by the neutrino oscillation experimental data, neutrinoless double beta decay, dark matter relic density, charged lepton flavor violation, electron-muon conversion, and provides essential means for efficient low-scale resonant leptogenesis. We have also shown that in the model considered for the analysis, the charged lepton flavor violating decays $\mu\to e\gamma$, $\mu\to eee$ as well as the electron-muon conversion processes get sizable rates, which are within the reach of sensitivity of the forthcoming experiments. Finally, we found that consistency with the measured value of the baryon asymmetry of the Universe requires that the entries of the Majorana submatrix corresponding to the LNV $\mu$ parameter acquire values in the keV to MeV range.

\section*{Acknowledgments}
This project has received funding and support from the European Union's Horizon 2020 research and innovation programme under the Marie Sk{\l }odowska-Curie grant agreement No.~860881 (H2020-MSCA-ITN-2019 HIDDeN) and from the Marie Skłodowska-Curie Staff Exchange  grant agreement No 101086085 ``ASYMMETRY''. 
A.E.C.H and S.K are supported by ANID-Chile FONDECYT 1210378, 1230160, ANID PIA/APOYO AFB220004, and Proyecto Milenio- ANID: ICN2019$\_$044.
TBM acknowledges ANID-Chile grant FONDECYT No. 3220454  for financial support.

\appendix
\section{\boldmath Form factors $\mu$-$e$ conversion and BR$(\mu \to eee)$} \label{formfactors}
In this appendix, we provide the analytical expressions for the form factors and the loop functions entering in the amplitudes of the muon-electron conversion process.
 
In the following, we provide the expressions for the radiative decay and $\ell_\beta \to 3 \ell_\alpha$ decays (the full expression for the most general case of the 3-body decay can be found in Ref.~\cite{Ilakovac:1994kj}), closely following the notation of Ref.~\cite{Abada:2021zcm}.
The rates for the radiative and three-body decays in the SM extended via
heavy RH neutrinos, are given by~\cite{Alonso:2012ji}
\begin{equation}
    \mathrm{BR}(\ell_\beta\to \ell_\alpha \gamma) \,= \frac{\alpha_w^3\,
      s_w^2}{256\,\pi^2}\,\frac{m_{\beta}^4}{M_W^4}\, \frac{m_{\beta}}{\Gamma_{\beta}}\, 
    \left|G_\gamma^{\beta \alpha} \right|^2\,, 
\end{equation}
\begin{align}
    \mathrm{BR}(\ell_\beta\to 3\ell_\alpha) =& \frac{\alpha_w^4}{24576\,\pi^3}\,\frac{m_{\beta}^4}{M_W^4}\, \frac{m_{\beta}}{\Gamma_{\beta}}\times\left\{2\left|\frac12F_\text{box}^{\beta 3\alpha} +F_Z^{\beta\alpha} - 2 s_w^2\,(F_Z^{\beta\alpha} - F_\gamma^{\beta\alpha})\right|^2 \right.  \nonumber\\ 
     &+ \left. 4 s_w^4\, |F_Z^{\beta\alpha} - F_\gamma^{\beta\alpha}|^2 + 16 s_w^2\,\mathrm{Re}\left[(F_Z^{\beta\alpha} - \frac12F_\text{box}^{\beta 3\alpha})\,G_\gamma^{\beta \alpha \ast}\right]\right.\nonumber\\ 
     &-\left. 48 s_w^4\,\mathrm{Re}\left[(F_Z^{\beta\alpha} - F_\gamma^{\beta\alpha})\,G_\gamma^{\beta\alpha \ast}\right] + 32 s_w^4\,|G_\gamma^{\beta\alpha}|^2\left[\log\frac{m_{\beta}^2}{m_{\alpha}^2} - \frac{11}{4}\right] \right\},
\end{align}
where $M_W$ is the $W$-boson mass, $\alpha_w = g_w^2/4\pi$ denotes the weak coupling, $s_w$ the sine of the weak mixing angle, and $m_{\beta}$ ($\Gamma_\beta$) the mass (total width) of the decaying charged lepton of flavor $\beta$. The form factors $G_\gamma^{\beta \alpha}$, $F_\gamma^{\beta \alpha}$, and $F_\text{box}^{\beta 3 \alpha}$ are given by~\cite{Ilakovac:1994kj, Alonso:2012ji}
\begin{align}
    G_\gamma^{\beta \alpha} &= \sum_{i =1}^{3 + n_S}
    \mathbb{U}_{\alpha i}^{\phantom{\ast}}\,\mathbb{U}_{\beta i}^\ast\,
    G_\gamma(x_i)\,,\label{eq:cLFV:FF:Ggamma} \\
     F_\gamma^{\beta \alpha} &= \sum_{i =1}^{3 + n_S}
    \mathbb{U}_{\alpha i}^{\phantom{\ast}}\,\mathbb{U}_{\beta i}^\ast
    \,F_\gamma(x_i)\,,\\ 
    F_Z^{\beta \alpha} &= \sum_{i,j =1}^{3 + n_S}
    \mathbb{U}_{\alpha i}^{\phantom{\ast}}\,\mathbb{U}_{\beta j}^\ast
    \left[\delta_{ij} \,F_Z(x_j) + 
    C_{ij}\, G_Z(x_i, x_j) + C_{ij}^\ast \,H_Z(x_i,
    x_j)\right], \label{eq:cLFV:FF:FZ}\\  
    F_\text{box}^{\beta 3 \alpha} &= \sum_{i,j = 1}^{3+n_S}
    \mathbb{U}_{\alpha i}^{\phantom{\ast}}\,\mathbb{U}_{\beta
      j}^\ast\left[\mathbb{U}_{\alpha i}^{\phantom{\ast}} \,\mathbb{U}_{\alpha
        j}^\ast\, G_\text{box}(x_i, x_j) - 2 \,\mathbb{U}_{\alpha
        i}^\ast \,\mathbb{U}_{\alpha j}^{\phantom{\ast}}\, F_\text{Xbox}(x_i, x_j)
      \right].\label{eq:cLFV:FF:Fbox}
    \end{align}
In the above expressions, the sums are understood to be taken over all neutral mass eigenstates. $C_{ij}$  is defined as 
\begin{equation}
    C_{ij} = \sum_{\rho = 1}^3 \mathbb{U}_{i\rho}^\dagger \,\mathbb{U}_{\rho j}^{\phantom\dagger }\,.
\end{equation}

The neutrinoless conversion rate is given by~\cite{Alonso:2012ji}   
\begin{equation}\label{eq:def:CRfull}
    \mathrm{CR}(\mu - e,\,\mathrm{N}) = \frac{2 G_F^2\,\alpha_w^2\,
      m_\mu^5}{(4\pi)^2\,\Gamma_\text{capt}}\left|4 V^{(p)}\left(2
    \widetilde F_u^{\mu e} + \widetilde F_d^{\mu e}\right) + 4 V^{(n)}\left(
    \widetilde F_u^{\mu e} + 2\widetilde F_d^{\mu e}\right)  + s_w^2
    \frac{G_\gamma^{\mu e}D}{2 e}\right|^2\,,  
\end{equation}
in which $\Gamma_\text{capt}$ denotes the capture rate for the nucleus N, with $D$, $V^{(p)}$ and $V^{(n)}$ corresponding to nuclear form factors (see Ref.~\cite{Kitano:2002mt}), and $e$ is the unit electric charge. The above form factors are given by~\cite{Alonso:2012ji, Ilakovac:1994kj} 
\begin{align}
    \widetilde F^{\mu e}_d &= -\frac{1}{3}s_w^2 F_\gamma^{\mu e} - F_Z^{\mu e}\left(\frac14 - \frac{1}{3}s_w^2 \right) + \frac14F^{\mu e dd}_\text{box}\ ,\\
    \widetilde F^{\mu e}_u &= \frac{2}{3}s_w^2 F_\gamma^{\mu e} + F_Z^{\mu e}\left(\frac14 - \frac{2}{3}s_w^2 \right) + \frac14F^{\mu e uu}_\text{box}\,,
\end{align}
to which one must add 
\begin{align}
     F_\text{box}^{\mu e uu} &= \sum_{i = 1}^{3 + n_S}\sum_{q_d = d, s, b} \mathbb{U}_{e i}^{\phantom{\ast}}\,\mathbb{U}_{\mu i}^\ast\, V_{u q_d}^{\phantom{\ast}}\,V_{u q_d}^\ast \:F_\text{box}(x_i, x_{q_d})\,, \label{eq:cLFV:FF:mueuu}\\
    F_\text{box}^{\mu e dd} &= \sum_{i = 1}^{3 + n_S}\sum_{q_u = u, c, t} \mathbb{U}_{e i}^{\phantom{\ast}}\,\mathbb{U}_{\mu i}^\ast\, V_{q_u d}^{\phantom{\ast}}\,V_{q_u d}^\ast \:F_\text{box}(x_i, x_{q_u})\,. \label{eq:cLFV:FF:muedd}    
\end{align}
Here, $x_{q} ={m_{q}^2}/{M_W^2}$ and $V$ is the Cabibbo-Kobayashi-Maskawa quark mixing matrix. The different loop functions (with arguments defined as $x_i ={m_{i}^2}/{M_W^2}$) are summarized below. 

\paragraph{Photon dipole and anapole functions} 
\begin{align}
    F_\gamma(x) &= \frac{7 x^3 - x^2 - 12x}{12(1-x)^3} - \frac{x^4 -
      10x^3 + 12x^2}{6(1-x)^4}\log x\,,\nonumber\\ 
    F_\gamma(x) &\xrightarrow[x\gg1]{} -\frac{7}{12} -
    \frac{1}{6}\log x\,,\nonumber\\ 
    F_\gamma(0) &= 0\,,\label{eqn:lfun:fgamma}\\
    G_\gamma(x) &= -\frac{x(2x^2 + 5x - 1)}{4(1-x)^3} -
    \frac{3x^3}{2(1-x)^4}\log x\,,\nonumber\\ 
    G_\gamma(x) &\xrightarrow[x\gg1]{} \frac12\,,\nonumber\\
    G_\gamma(0) &= 0\,.\label{eqn:lfun:ggamma}
\end{align}

\paragraph{\boldmath $Z$-penguin: two- and three-point functions}
\begin{align}
    F_Z(x) &= -\frac{5 x}{2(1 - x)} - \frac{5x^2}{2(1-x)^2}\log x\,,\nonumber\\
    F_Z(x) &\xrightarrow[x\gg 1]{} \frac{5}{2} - \frac{5}{2}\log
    x\,,\nonumber\\ 
    F_Z(0) &= 0\,,\label{eqn:lfun:fz}
\end{align}
\begin{align}
    G_Z(x,y) &= -\frac{1}{2(x-y)}\left[\frac{x^2(1-y)}{1-x}\log x - \frac{y^2(1-x)}{1-y}\log y \right],\nonumber\\ 
    G_Z(x, x) &= -\frac{x}{2} - \frac{x\log x}{1-x}\,,\nonumber\\
    G_Z(0,x) &= -\frac{x\log x}{2(1-x)}\,,\nonumber\\
    G_Z(0,x)&\xrightarrow[x\gg 1]{} \frac12\log x\,,\nonumber\\
    G_Z(0,0) &= 0\,,\label{eqn:lfun:gz}\\
    H_Z(x,y) &= \frac{\sqrt{xy}}{4(x-y)}\left[\frac{x^2 - 4x}{1 - x}\log x - \frac{y^2 - 4y}{1 - y}\log y\ \right],\nonumber\\ 
    H_Z(x,x) &= \frac{(3 - x)(1-x) - 3}{4(1-x)} - \frac{x^3 - 2x^2 + 4x}{4(1-x)^2}\log x\,,\nonumber\\ 
    H_Z(0,x) &= 0\,.\label{eqn:lfun:hz}
\end{align}

\paragraph{Box loop-functions}
\begin{align}
    F_\text{box}(x,y) &= \frac{1}{x-y}\left\{\left(4 + \frac{xy}{4}\right)\left[\frac{1}{1-x} + \frac{x^2}{(1-x)^2} \log x - \frac{1}{1-y} - \frac{y^2}{(1-y)^2}\log y\right]\right.\nonumber\\  
    &\phantom{=} \left. -2xy\left[\frac{1}{1-x} + \frac{x}{(1-x)^2} \log x - \frac{1}{1-y} - \frac{y}{(1-y)^2}\log y \right]\right\},\nonumber\\ 
    F_\text{box}(x,x) &= -\frac{1}{4(1-x)^3}\left[x^4 - 16x^3 + 31x^2 - 16 + 2x\left(3x^2 + 4x - 16\right)\log x\right],\nonumber\\ 
    F_\text{box}(0,x) &= \frac{4}{1 - x} + \frac{4x}{(1-x)^2}\log x\,,\nonumber\\ 
    F_\text{box}(0,x)&\xrightarrow[x\gg 1]{} 0\,,\nonumber\\
    F_\text{box}(0,0) &= 4\,,\label{eqn:lfun:fbox}\\
    F_\text{Xbox}(x,y) &= -\frac{1}{x-y}\left\{\left(1 + \frac{xy}{4} \right)\left[\frac{1}{1-x} + \frac{x^2}{(1-x)^2} \log x - \frac{1}{1-y} - \frac{y^2}{(1-y)^2}\log
      y\right]\right.\nonumber\\  
    &\phantom{=} \left. -2xy\left[\frac{1}{1-x} + \frac{x}{(1-x)^2} \log x - \frac{1}{1-y} - \frac{y}{(1-y)^2}\log y \right]\right\},\nonumber\\ 
    F_\text{Xbox}(x,x) &= \frac{x^4 - 16x^3 + 19x^2 - 4}{4(1-x)^3} + \frac{3x^3 + 4x^2 - 4x}{2(1-x)^3}\log x\,,\nonumber\\ 
    F_\text{Xbox}(0,x) &= -\frac{1}{1-x} - \frac{x}{(1 - x)^2}\log x\,,\nonumber\\ 
    F_\text{Xbox}(0,x)&\xrightarrow[x\gg 1]{} 0\,,\nonumber\\
    F_\text{Xbox}(0,0) &= -1\,,\label{eqn:lfun:fxbox}\\
    G_\text{box}(x,y) &= -\frac{\sqrt{xy}}{x-y}\left\{(4 + xy)\left[\frac{1}{1-x} + \frac{x}{(1-x)^2} \log x - \frac{1}{1-y} - \frac{y}{(1-y)^2}\log y\right]\right.\nonumber\\  
    &\phantom{=} \left. -2\left[\frac{1}{1-x} + \frac{x^2}{(1-x)^2} \log x - \frac{1}{1-y} - \frac{y^2}{(1-y)^2}\log y \right]\right\},\nonumber\\ 
    G_\text{box}(x,x) &= \frac{2x^4 - 4x^3 + 8x^2 - 6x}{(1-x)^3} - \frac{x^4 + x^3 + 4x}{(1-x)^3}\log x\,,\nonumber\\ 
    G_\text{box}(0,x) &= 0\,.\label{eqn:lfun:gbox}
\end{align}

\bibliographystyle{utphys}
\bibliography{Refs}
\end{document}